\documentclass[linenumber]{aastex}
\usepackage[utf8]{inputenc}
\usepackage{natbib}
\usepackage{CJKutf8}
\usepackage{gensymb}
\usepackage{float}
\usepackage{amsmath}
\usepackage[mathlines]{lineno}
\usepackage{appendix}
\usepackage{subfigure} 

\begin{document}
\begin{CJK}{UTF8}{gbsn}

\title{The First Photometric Analysis of Two Low Mass Ratio Contact Binary Systems In TESS Survey}
\author{Qiyuan Cheng}
\affiliation{Yunnan Observatories, Chinese Academy of Sciences, 396 YangFangWang, Guandu District, Kunming, 650216, Peopleʼs Republic of China}
\affiliation{School of Astronomy and Space Science, University of Chinese Academy of Sciences, Beijing, 100049, China}

\author[0000-0003-4829-6245]{Jianping Xiong}
\affiliation{Yunnan Observatories, Chinese Academy of Sciences, 396 YangFangWang, Guandu District, Kunming, 650216, Peopleʼs Republic of China}
\affiliation{Key Laboratory for Structure and Evolution of Celestial Objects, Chinese Academy of Sciences, P.O. Box 110, Kunming 650216, People's Republic of China}
\affiliation{International Centre of Supernovae, Yunnan Key Laboratory, Kunming 650216, People's Republic of China}

\author{Xu Ding}
\affiliation{Yunnan Observatories, Chinese Academy of Sciences, 396 YangFangWang, Guandu District, Kunming, 650216, Peopleʼs Republic of China}

\author{Kaifan Ji}
\affiliation{Yunnan Observatories, Chinese Academy of Sciences, 396 YangFangWang, Guandu District, Kunming, 650216, Peopleʼs Republic of China}

\author{Jiao Li}
\affiliation{Key Laboratory of Space Astronomy and Technology, National Astronomical Observatories, CAS, Beijing 100101, People’s Republic of China}

\author{Chao Liu}
\affiliation{Key Laboratory of Space Astronomy and Technology, National Astronomical Observatories, CAS, Beijing 100101, People’s Republic of China}
\affiliation{Institute for Frontiers in Astronomy and Astrophysics, Beijing Normal University, Beijing, 102206, China}
\affiliation{School of Astronomy and Space Science, University of Chinese Academy of Sciences, Beijing, 100049, China}

\author{Jiangdan Li}
\affiliation{Yunnan Observatories, Chinese Academy of Sciences, 396 YangFangWang, Guandu District, Kunming, 650216, Peopleʼs Republic of China}

\author{Jingxiao Luo}
\affiliation{Yunnan Observatories, Chinese Academy of Sciences, 396 YangFangWang, Guandu District, Kunming, 650216, Peopleʼs Republic of China}
\affiliation{School of Astronomy and Space Science, University of Chinese Academy of Sciences, Beijing, 100049, China}

\author[0000-0001-8550-0095]{Xin Lyu}
\affiliation{National Astronomical Observatories, Chinese Academy of Sciences, Beijing 100101, China}
\affiliation{School of Astronomy and Space Science, University of Chinese Academy of Sciences, Beijing, 100049, China}
\affiliation{Key Laboratory of Radio Astronomy and Technolgoy, Chinese Academy of Sciences, A20 Datun Road, Chaoyang District, Beijing, 100101, P.R. China}

\author[0000-0001-9204-7778]{Zhanwen Han}
\affiliation{Yunnan Observatories, Chinese Academy of Sciences, 396 YangFangWang, Guandu District, Kunming, 650216, Peopleʼs Republic of China}
\affiliation{Key Laboratory for Structure and Evolution of Celestial Objects, Chinese Academy of Sciences, P.O. Box 110, Kunming 650216, People's Republic of China}
\affiliation{International Centre of Supernovae, Yunnan Key Laboratory, Kunming 650216, People's Republic of China}

\author[0000-0001-5284-8001]{Xuefei Chen}
\affiliation{Yunnan Observatories, Chinese Academy of Sciences, 396 YangFangWang, Guandu District, Kunming, 650216, Peopleʼs Republic of China}
\affiliation{Key Laboratory for Structure and Evolution of Celestial Objects, Chinese Academy of Sciences, P.O. Box 110, Kunming 650216, People's Republic of China}
\affiliation{International Centre of Supernovae, Yunnan Key Laboratory, Kunming 650216, People's Republic of China}

\correspondingauthor{Xuefei Chen.}
\email{cxf@ynao.ac.cn}


\begin{abstract}
Low mass-ratio ($q$) contact binary systems are progenitors of stellar mergers such as \textbf{blue straggles (BS)} or fast-rotating FK Com stars. In this study, we present the first light curve analysis of two newly identified low mass-ratio contact binary systems, TIC 55007847 and TIC 63597006, that are identified from TESS. Both stars are classified as A-subtype contact binaries. We obtained the precise orbit periods for the two objects by using the O-C method, i.e. P=0.6117108 d for TIC 55007847 and P=0.7008995 d for TIC 63597006, respectively, and found an obvious periodic signal in the O-C curve of TIC 63597006. We suggest that the periodic signal comes from a third body.
We further use the Markov Chain Monte Carlo (MCMC) method with PHOEBE to derive the photometric solutions for the two binaries. The photometric solution for this object shows that the contribution of the third body is about 6\%. Our analysis revealed that TIC 55007847 has an extremely low mass ratio of $q=0.08$. 
By calculating the ratio of spin angular momentum to the orbital angular momentum $J_{\rm s}$/$J_{\rm o}$, we found that TIC 55007847 is very close to the instability threshold with $J_{\rm s}$/$J_{\rm o}$ = 0.31, indicating that it may merge into a single, fast-rotating star in the future. 
For TIC 63597006, $q=0.14$ and $J_{\rm s}$/$J_{\rm o}=0.15$.
This object is in a relatively stable evolutionary status at present.

\end{abstract}

\section{Introduction}

In contact binaries, both of the components overfill their inner critical equipotential surface, forming a common envelope \citep{1959cbs..book.....K}.
There are two sub-types of contact binaries i.e. A subtype and W subtype,  which represent different evolutionary statuses of contact binaries. Typically, for the A subtype, the more massive component is hotter than the less massive one, representing the early status of contact binaries, while for the W subtype, the more massive star is cooler than the secondary one, representing relatively late evolutionary status \citep{1970VA.....12..217B,1977VA.....21..359B}. 
 
Contact binaries play a significant role in the study of binary evolution, 
including mass and energy \textbf{transfer} in the common envelope. 
Meanwhile, contact binaries and their mergers may appear as some peculiar stars such as blue stragglers and fast-rotating FK Com stars \citep{1995ApJ...444L..41R,2021RAA....21..235W} .
The lower limit of mass ratio for contact binaries is a long-term question 
that has attracted much attention in the last decades.
 
Due to the Darwin instability \citep{1966ARA&A...4...35H,1980A&A....92..167H}, contact binaries exhibit a low mass ratio cutoff ($q_{\rm min}$) \citep{1970VA.....12..217B,1976ApJ...209..829W}. 
When the mass ratio $q$ (the less massive one/the more massive one) of a contact binary falls below the $q_{\rm min}$, the tidal instability leads to the merger of two components.
The statistical research shows that the value of $q_{\rm min}$ may be as low as $0.044(\pm0.007)$ \citep{2015AJ....150...69Y}. 
A recent study of \citet{2023A&A...672A.176P} indicates that
the value of $q_{\rm min}$ may be different for various types of contact binaries. For late-type contact binaries, $q_{\rm min}=0.087^{+0.024}_{-0.015}$ for those with periods longer than $0.3$ days and $q_{\rm min}=0.0246^{+0.029}_{-0.046}$ for those with shorter periods, 
while for early-type ones, $q_{\rm min}=0.030^{+0.018}_{-0.022}$. Furthermore, it has been shown that the mergers are accompanied by detectable transients. For example, V1309 Sco is a red nova eruption that is believed to yield from the merging of a contact binary \citep{2011A&A...528A.114T, 2011cxo..prop.3317R}. KIC 9832227 is identified as a progenitor candidate for an imminent red nova ($q=0.228 \pm 0.003$, $P=0.45796  \rm d$), with rapid period decreasing rate $\dot P $ $> 1\times10^{-8}$ under an exponentially decaying function $(P \propto exp(\frac{1}{t-t_{0}})$, $t_{0}$ is the time of merging and $t$ is the time of observation) used for V1309 Sco \citep{2017ApJ...840....1M}.
Searching for contact binary systems with low or extremely low mass ratios is therefore an important research topic. For example, \citet{2022yCat..51640202L} has proposed a catalog of ten candidates of low mass ratio contact binaries. \citet{2022MNRAS.512.1244C} has analyzed 30 candidates from the Catalina Sky Survey. Furthermore, \citet{2023A&A...672A.176P} has provided 300 candidates from the Kepler catalog.

TESS (Transiting Exoplanet Survey Satellite) 
provides an excellent opportunity to discover variable stars as well as \textbf{eclipsing} binaries (EBs), although it is designed for searching for exoplanets \citep{2015JATIS...1a4003R}.
For example, \citet{2021AcASn..62...39T} discovered 4625 variable stars from about 20000 high-\textbf{quality} light curves. \citet{2021A&A...652A.120I} has provided confirmation of 3155 candidates of O-type, B-type, and A-type EBs, 
and \citet{2022ApJS..258...16P} collected 4584 EBs in sectors 1-26 of short-cadence observations. 
Meanwhile, \citet{2021PASJ...73..786D} has provided a helpful machine learning method to acquire parameters of contact binary including inclination, mass ratio, fill-out factor, and temperature ratio. 
In this paper, we identify $2$ contact binaries with low-mass ratios using the method of \citet{2021PASJ...73..786D} and investigate them in detail. The IDs of the two binaries are TIC 55007847 and TIC 63597006 \footnote{All of the data presented in this article were obtained from the Mikulski Archive for Space Telescopes (MAST) at the Space Telescope Science Institute. The specific observations analyzed can be accessed via \dataset[DOI: 10.17909/h37d-c176]{https://doi.org/10.17909/h37d-c176}}, with mass ratios of $q=0.0806$ and $q=0.141$, respectively. The photometric solutions as well as absolute parameters (luminosity, radii, masses, surface gravity, and period), and evolutionary statuses are calculated for the two objects for the first time.

The paper is structured as follows. The orbital periodic correction is described in Section~\ref{sec:Orbital}. We present the measurement of accurate photometric parameters in Section~\ref{sec:lightcurve}. The estimation of stellar parameters is shown in Section~\ref{sec:method}. The merge status is indicated in Section~\ref{sec:dis}. Finally, we summarize in Section~\ref{sec:con}.

\section{Orbital periods Analysis}\label{sec:Orbital}

\subsection{The data} \label{sec:data}
\begin{figure*}
    \centering
    \includegraphics[width=0.85\textwidth]{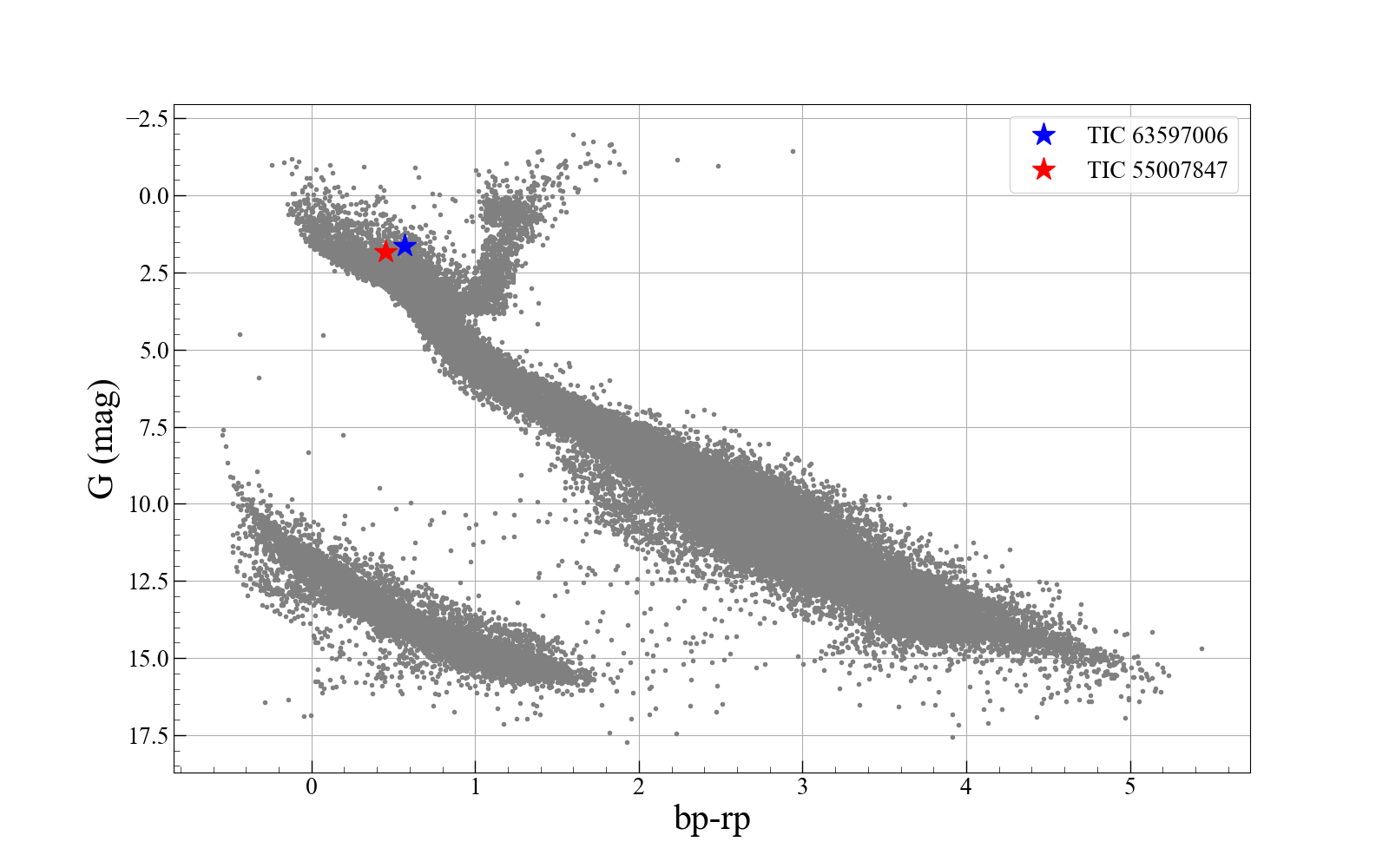}
    \caption{The Hertzsprung-Russell diagram of TIC 55007847 and TIC 63597006. The red star is the TIC 55007847. The blue star is the TIC 63597006. \textbf{The gray dots represent the stars selected from Gaia DR3 using the selection method presented by \citet{2018A&A...616A...2L}, which have a distance of less than 100 pc.}}
    \label{fig:figure 1}
\end{figure*}

The observed intensities on multi-band photometry of two targets and locations are summarized in \textbf{Table \ref{tbl: Table 1}}. TIC 55007847 has 13689 photometric measurements by TESS in a time span of 24.162 days. Similarly, TIC 63597006 has 18348 photometric measurements in a time span of 27.838 days. The single exposure time of each target is two minutes. \textbf{Fig \ref{fig:figure 1}} 
shows the positions of the two targets on the Hertzsprung-Russell diagram based on Gaia DR3, indicating they are likely F-type stars. 

\begin{deluxetable*}{lcc} \label{tbl: Table 1}
\tablewidth{40pt} 
\tablecaption{Basic information of two low mass ratio conatact binaries.}
\tablehead{\colhead{Basic Infomation} & \colhead{TIC 55007847} & \colhead{TIC 63597006}
} 
\startdata 
{$\mathrm{R.A.}\ (\mathrm{J2000})$} & {$10^\mathrm{h}44^\mathrm{m}21^\mathrm{s}.98$} & {$18^\mathrm{h}04^\mathrm{m}32^\mathrm{s}.82$} \\
{$\mathrm{Decl.}\ (\mathrm{J2000})$} & {$-7^\circ11^\prime08^{\prime \prime}.9$} & {$-42^\circ13^\prime24^{\prime \prime}.3$} \\
{B (Hipparcos)} & {$11.90$ mag} & {$8.83$ mag} \\
{V (Hipparcos)} & {$11.39$ mag} & {$8.37$ mag} \\
{G (Gaia EDR3)} & {$11.49$ mag} & {$8.20$ mag} \\
{J (2MASS)} & {$10.95$ mag} & {$7.61$ mag} \\
{H (2MASS)} & {$10.80$ mag} & {$7.44$ mag} \\
{K (2MASS)} & {$10.76$ mag} & {$7.34$ mag} \\
\hline
\enddata
\end{deluxetable*}

\subsection{The O-C analysis}

For the light curves, we first obtain the orbital periods $P_{\rm 0}$ from the LombScargle algorithm \citep{2009A&A...496..577Z} and correct them utilizing the O-C method as introduced below. If we have an orbital period $P_0$, the primary minima of a light curve are written as \citep{1950BAN....11..217O}:

\begin{equation}\label{eq:eq1}
    {\rm Min}\ I= T_{\mathrm{0}} + P_{\rm 0} \times N,
\end{equation}

where $T_{\rm 0}$ is the time of zero epoch and $N$ is the epoch number. 
The O-C curve is defined as the difference of the primary minima 
between the observation (O) and the calculation (C) \citep{2019OEJV..197.....K}, that is, 
\begin{equation}\label{eq:eq2}
O-C =\Delta T_0 + \Delta P \times N + \tau, 
\end{equation}
where $\Delta T_{\rm 0}$ represents the deviation of the time of zero epoch, $\Delta P$ is the deviation of the period between the period from LombScargle with the true period, $\tau$ is related to more complex periodic variations
such as the third body light time orbital effect or periodic magnetic activity \citep{2019OEJV..197.....K}. 

Firstly, we extract all of the primary minima of light curves in Barycentric Dynamical Time (BJD), and define the first minimum as the time of zero ephemeris $T_{\rm 0}$, 
and calculate the primary minima using \textbf{Eq (\ref{eq:eq1})} and $P_0$. 
The O-C values are obtained and are fitted by \textbf{Eq (\ref{eq:eq2})} using the Markov Chain Monte Carlo (MCMC) method,
which will give the values of $\Delta T_{\rm 0}$ and $\Delta P$, consequently.
We then may correct the period to be $P_0+\Delta P$. 
\textbf{Table \ref{Tbl: Table 2}} summarizes the observations and O-C values for the two objects.

\subsection{Results}
\begin{figure*}[ht] 
    \centering
    \includegraphics[width=0.85\textwidth]{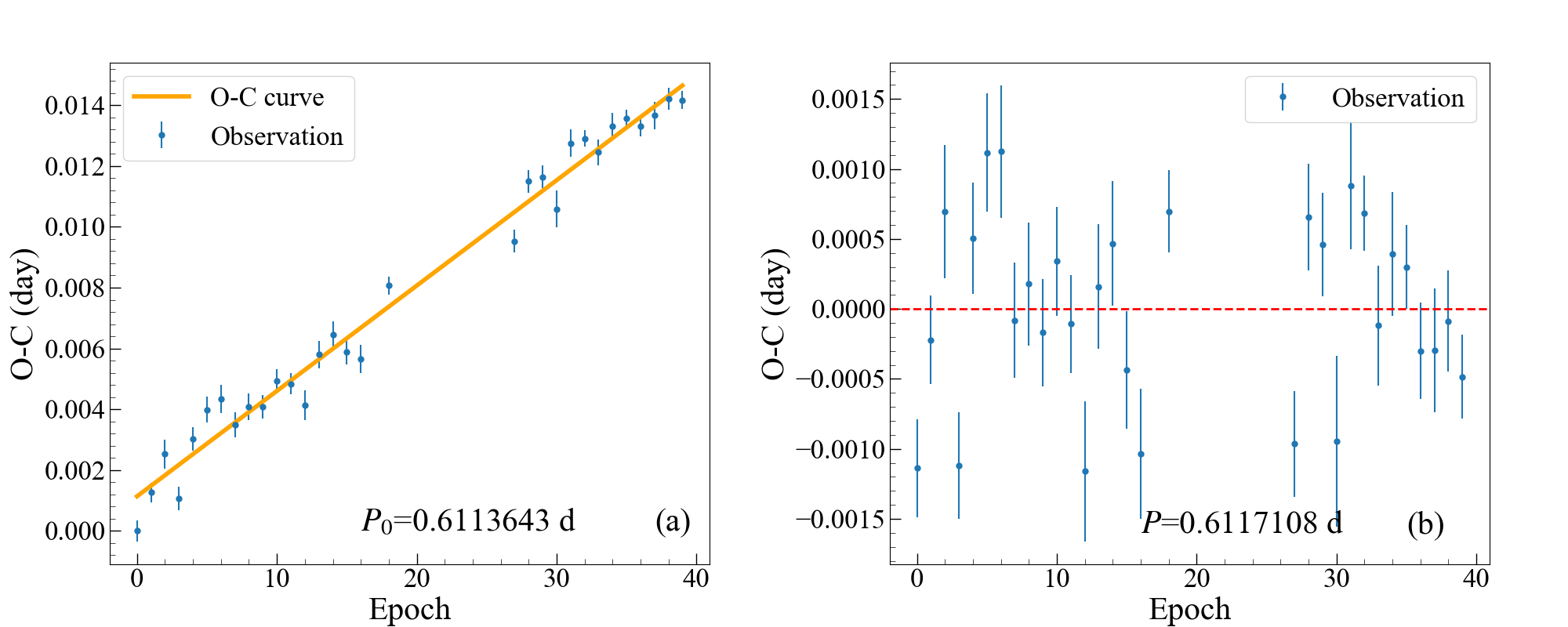}
    \caption{The O-C diagram for the object TIC 55007847. Panel (a) shows the result from the LombScargle method ($P_0$=0.6113643 d), while panel (b) is for the corrected period ($P$=0.6117108 d).}
    \label{fig:figure 2}
\end{figure*}

For TIC 55007847, TESS has provided approximately 24 days of observation, 
$T_{\rm 0}$ = 2255.95697 d, and the LombScargle method gives an orbital period of 0.6113643 d. \textbf{Fig \ref{fig:figure 2} (a)} shows the O-C diagram,
where the blue dots are the primary minima with error bars and the solid line is the fitted O-C curve as below:

\begin{equation} \label{eq:eq3}
    O-C =  {3.46481}^{+0.05507}_{-0.05622} \times10^{-4} \times N + {1.13642}^{+0.12641}_{-0.12308} \times10^{-3},
\end{equation}

We see in the figure an almost linear relationship between the epoch and the O-C values, indicating that the orbital period is varying or that the true orbital period is different from that of the adopted. Since the time span for this object is only about 24 days, it is impossible to see the variation of the period. 
The linear relation is therefore due to the deviation of the true period from the adopted one.
According \textbf{Eq (\ref{eq:eq3})}, we have $\Delta P=3.46481\times10^{-4}$ d. 
The true period of TIC 55007847 is therefore 0.6117108 d.

Based on the corrected period, we recalculate the primary minima and obtain new O-C values, as shown in \textbf{Fig \ref{fig:figure 2} (b)}. Now we see that the O-C values are randomly distributed around zero, suggesting a good solution for the period in this way.

\begin{deluxetable}{cccc|cccc} \label{Tbl: Table 2}
\tablewidth{40pt} 
\tablecaption{The primary minima information for TIC 55007847 and TIC 63597006. The primary minima are extracted by the quadratic curve fit. The first column of each target is the time of primary minimum in BJDs-2457000 d. The $\Delta T$ is the uncertainty of each primary minimum in the second column. The third column shows the epoch number of each primary minimum, and the forth column is the value of O-C. The periods ($P_0$) for TIC 55007847 and TIC 63597006 from LombScargle are 0.6113643 d and 0.7005181 d, respectively. The time of zero epochs ($T_0$) for TIC 55007847 and TIC 63597006 is 2255.95697 d and 2362.14929 d, respectively.}
\tablehead{
\multicolumn{4}{c}{TIC 55007847} & \multicolumn{4}{c}{TIC 63597006} \\
\hline
\colhead{BJDs - 2457000 (d)} & \colhead{$\Delta T$ (d)}& \colhead{Epoch} & \colhead{O-C (d)} & \colhead{BJDs - 2457000 (d)} & \colhead{$\Delta T$ (d)}& \colhead{Epoch} & \colhead{O-C (d)}\\
} 
\startdata 
2255.95697 & 0.00035 & 0 & 0.00000 & 2362.14929 & 0.00019 & 0 & 0.00000 \\
2256.56960 & 0.00032 & 1 & 1.39848 & 2362.85091 & 0.00024 & 1 & 1.58126 \\
2257.18222 & 0.00048 & 2 & 2.79697 & 2363.55172 & 0.00022 & 2 & 1.99889 \\
2257.79212 & 0.00038 & 3 & 0.26818 & 2364.25303 & 0.00022 & 3 & 3.14379 \\
2258.40545 & 0.00040 & 4 & 2.68484 & 2364.95303 & 0.00022 & 4 & 2.39779 \\
2259.01778 & 0.00042 & 5 & 3.64696 & 2365.65434 & 0.00021 & 5 & 3.54269 \\
2259.62949 & 0.00047 & 6 & 3.73635 & 2366.35434 & 0.00021 & 6 & 2.79668 \\
2260.24000 & 0.00041 & 7 & 2.08029 & 2367.05576 & 0.00026 & 7 & 4.08704 \\
2260.85197 & 0.00044 & 8 & 2.53332 & 2367.75576 & 0.00026 & 8 & 3.34103 \\
2261.46333 & 0.00038 & 9 & 2.11362 & 2368.45697 & 0.00025 & 9 & 4.34047 \\
2262.07556 & 0.00039 & 10 & 2.93028 & 2369.15818 & 0.00025 & 10 & 5.33992 \\
2262.68682 & 0.00035 & 11 & 2.36513 & 2369.85939 & 0.00025 & 11 & 6.33937 \\
2263.29747 & 0.00050 & 12 & 0.92725 & 2370.56061 & 0.00025 & 12 & 7.33882 \\
2263.91051 & 0.00044 & 13 & 2.90755 & 2371.26061 & 0.00026 & 13 & 6.59281 \\
2264.52253 & 0.00044 & 14 & 3.43330 & 2371.96182 & 0.00026 & 14 & 7.59225 \\
2265.13333 & 0.00042 & 15 & 2.21360 & 2372.66303 & 0.00026 & 15 & 8.59170 \\
2265.74444 & 0.00047 & 16 & 1.43027 & 2373.36424 & 0.00026 & 16 & 9.59115 \\
2266.96960 & 0.00029 & 18 & 4.08178 & 2374.06545 & 0.00026 & 17 & 10.59059 \\
2272.47333 & 0.00038 & 27 & 2.41358 & 2374.76545 & 0.00027 & 18 & 9.84459 \\
2273.08667 & 0.00038 & 28 & 4.83024 & 2376.16788 & 0.00029 & 20 & 11.84348 \\
2273.69818 & 0.00037 & 29 & 4.62872 & 2376.86818 & 0.00025 & 21 & 11.53384 \\
2274.30848 & 0.00061 & 30 & 2.68175 & 2377.56939 & 0.00025 & 22 & 12.53328 \\
2274.92202 & 0.00045 & 31 & 5.38932 & 2378.27051 & 0.00036 & 23 & 13.38727 \\
2275.53354 & 0.00027 & 32 & 5.18781 & 2378.97152 & 0.00036 & 24 & 14.09581 \\
2276.14444 & 0.00043 & 33 & 4.11356 & 2379.67253 & 0.00036 & 25 & 14.80435 \\
2276.75667 & 0.00044 & 34 & 4.93023 & 2380.37354 & 0.00037 & 26 & 15.51289 \\
2277.36828 & 0.00030 & 35 & 4.87416 & 2381.07455 & 0.00039 & 27 & 16.22142 \\
2277.97939 & 0.00034 & 36 & 4.09083 & 2381.77556 & 0.00039 & 28 & 16.92996 \\
2278.59111 & 0.00044 & 37 & 4.18022 & 2382.47576 & 0.00026 & 29 & 16.47486 \\
2279.20303 & 0.00036 & 38 & 4.56052 & 2383.17667 & 0.00029 & 30 & 17.03795 \\
2279.81434 & 0.00030 & 39 & 4.06809 & 2383.87778 & 0.00030 & 31 & 17.89194 \\
{...} & {...} & {...} & {...} & 2384.57889 & 0.00030 & 32 & 18.74593 \\
{...} & {...} & {...} & {...} & 2385.27848 & 0.00035 & 33 & 17.41810 \\
{...} & {...} & {...} & {...} & 2385.97949 & 0.00036 & 34 & 18.12664 \\
{...} & {...} & {...} & {...} & 2386.68061 & 0.00025 & 35 & 18.98063 \\
{...} & {...} & {...} & {...} & 2387.38111 & 0.00032 & 36 & 18.96190 \\
{...} & {...} & {...} & {...} & 2389.48444 & 0.00035 & 39 & 21.52388 \\
\enddata
\end{deluxetable}

In a similar way, we derive the orbital period for TIC 63597006. For this object, $P_0$=0.7005181 d and $T_{\rm 0}$ = 2362.14929 d.
The O-C line is fitted by

\begin{equation}\label{eq:eq4}
    O-C = {3.81475}^{+0.04108}_{-0.04166} \times10^{-4} \times N + {2.64547}^{+0.08508\times10^{-5}}_{-0.08508\times10^{-5}} \times10^{-4},
\end{equation}
which gives $\Delta P=3.81475\times10^{-4}$ d. The true period is therefore corrected to be 0.7008995 d. 

\begin{figure}[ht]
    \centering
    \includegraphics[width=0.85\textwidth]{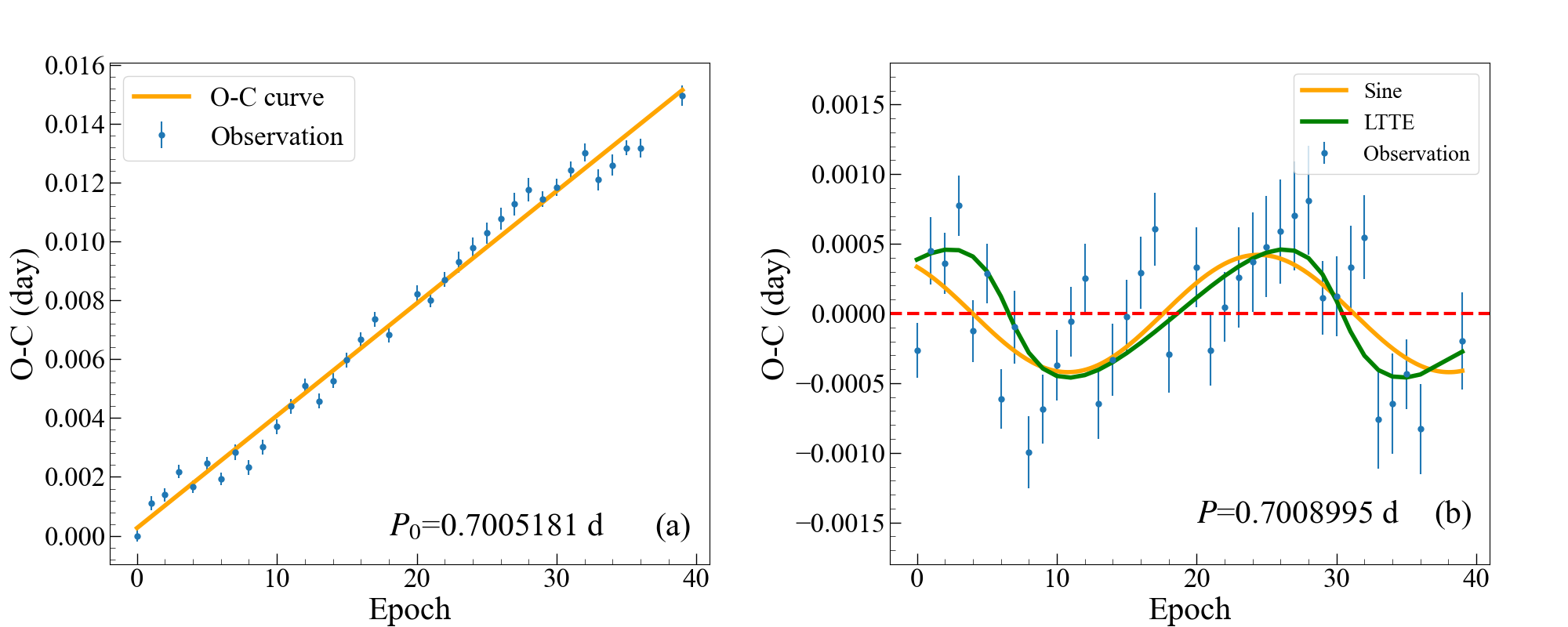}
    \caption{The O-C diagram for the object TIC 63597006. 
    Panel (a) shows the result from the LombScargle method ($P_0$=0.7005181 d), while panel (b) is for the corrected period ($P$=0.7008995 d). We see an obvious periodic signal in the panel (b) which could be caused by the orbital motion from the third body or the magnetic activities. The green solid line shows the result of LTTE and the orange solid line shows the sine function fitting. They represent the affection of the third body and magnetic activities respectively.}
    \label{fig:figure 3}
\end{figure}

The result is shown in \textbf{Fig \ref{fig:figure 3}}. Different from that of TIC 55007847, we can see an obvious periodic signal in the \textbf{Fig \ref{fig:figure 3} (b).} 
Such a periodic signal ($\tau$) of the O-C curve is possibly from the orbital motion of a third body \citep{1952ApJ...116..211I,2006AJ....131.2986P} or stellar magnetic activities, which can be fitted by the light travel time effect (LTTE) or the sine function, respectively \citep{2020ApJ...901..169L}. The LTTE can be write as \citep{1952ApJ...116..211I}:

\begin{equation} \label{eq:eq5}
\tau = K\frac{1}{\sqrt{1-e^2\ {\rm cos}^2\omega}}\left[ \frac{1-e^2}{1+e\ {\rm cos}\nu}{\rm sin}(\nu+\omega)+e\ {\rm sin}(\omega)\right],
\end{equation}
and 
\begin{equation} \label{eq:eq6}
K = \frac{1}{2}(\tau_{\rm max}-\tau_{\rm min})=\frac{a\ {\rm sin}i}{c}\sqrt{1-e^2\ {\rm cos}^2\omega},
\end{equation}
 
where $a$, $i$, $e$, $\nu$, and $\omega$ are the semi-major axis, inclination, eccentricity, true anomaly, and longitude of the periastron of orbital of the third body, respectively, and $c$ is the speed of light. 
The term of $a\ {\rm sin}i/c$ is the amplitude of LTTE. 
The true anomaly can be obtained by the following equations:

\begin{equation}\label{eq:eq7}
    {\rm tan}\frac{1}{2}E=\left( \frac{1-e}{1+e}\right)^{1/2} {\rm tan}\frac{1}{2} \nu, {\rm  and }
    \frac{t-T}{P} = \frac{M}{2\pi} = \frac{E-e\ {\rm sin}E}{2\pi},
\end{equation}

where $M$ is the mean anomaly, $T$ and $P$ are the zero epoch and the orbital period of the third body. The green solid line in \textbf{Fig \ref{fig:figure 3} (b)} shows the fitting result of LTTE. We have $a\ {\rm sin}i/c = 0.000519$ d, $e=0.4$, and $P=17.71786$ d from the fitting. 

We further made a sine function fitting to investigate the effect of  magnetic activities \citep{1992ApJ...385..621A,2020ApJ...901..169L}:

\begin{equation}\label{eq:eq9}
    \tau = A {\rm sin}\left( \frac{2\pi P N}{P_{\rm out}} + \phi\right),
\end{equation}
in which $A$ is the amplitude of the cyclic oscillation, $P_{\rm out}$ represents the period of an extra periodic signal and $\phi$ is the phase of epoch. The orange solid line in \textbf{Fig \ref{fig:figure 3} (b)} shows the fitting result, with $A=0.0004$ d, $P_{\rm out}=19.0704$ d and $\phi=2.22180$.

To analyze whether the periodic signal arises from magnetic activity or a third light, we employ the Bayesian method for assessment. According to Bayesian inference, the evidence can be written as:

\begin{equation}
    p(y)=\int p(y|\theta_{\rm i})p(\theta_{\rm i})d\theta_{\rm i},
\end{equation}
where the $p(y)$ is the evidence in the posterior possibility, $p(y|\theta_{\rm i})$ is the likelihood, and $p(\theta_{\rm i})$ is the prior possibility. We assume that the prior is uniform and the likelihood is Gaussian. The uncertainties of each primary minimum (the third column in \textbf{Table \ref{Tbl: Table 2}}) are adopted as the uncertainties in Gaussian likelihood. \textbf{Based on the specified fitting formula, one sampling in MCMC yields a fitting curve}. For each \textbf{fitting curve}, the likelihood of the line is the mean value of the likelihood of its sampling points. Because the uniform prior is used, the normalization constant is \textbf{the sum of the possibility of each} \textbf{fitting curve}. Then the evidence $p$ of the two effects and obtain $p_{\rm LTTE} = 0.5731$ and $p_{\rm m} = 0.2023$, indicating that the periodic signal is more likely from the third body. We also calculate the $\chi ^2$ of two physical pictures. \textbf{The ${\chi}^2_\mathrm{LTTE}=0.0272$ and \textbf{${\chi}^2_\mathrm{m}=0.1374$}} also show that the third body is a more reasonable explanation. Meanwhile, it is suggested that magnetic activities likely lead to the O'Connell effect \citep{1951PRCO....2...85O} 
which has not been seen in the light curve of TIC 63597006 (see next section), as is consistent with our result here.

\section{Photometric Solutions} \label{sec:lightcurve}
\subsection{Fitting Stategy}
We use the MCMC \citep{2013PASP..125..306F} approach with the PHOEBE (Physics Of Eclipsing Binaries) software (version 2.3.58) \citep{2016ApJS..227...29P} to obtain the photometric solutions for the two objects. we fold the light curves in the phase-magnitude diagram according to the periods obtained in the O-C section. The common free parameters are inclination ($incl$), mass ratio ($q$), fill-out factor ($f$), and temperature ratio of the secondary star over the primary star ($T_2/T_1$). The gravity-darkening coefficient and bolometric albedo are set as $A_1 = A_2 = 0.5$ \citep{1969AcA....19..245R} and $g_1 = g_2 = 0.32$ \citep{1967ZA.....65...89L}. We run the model with 30 walkers, each of which takes 2000 steps. The peak values and standard deviations of the posterior distribution yield are seen as the best-fit parameters and their uncertainties.

\subsection{TIC 55007847}
For TIC 55007847, we set the $T_1 = 7141\ \rm K$ as the effective temperature obtained from the Large Sky Area Multi-Object Fiber Spectroscopic Telescope (LAMOST) low-resolution spectrum.

\begin{figure*}
    \centering
    \includegraphics[width=0.85\textwidth]{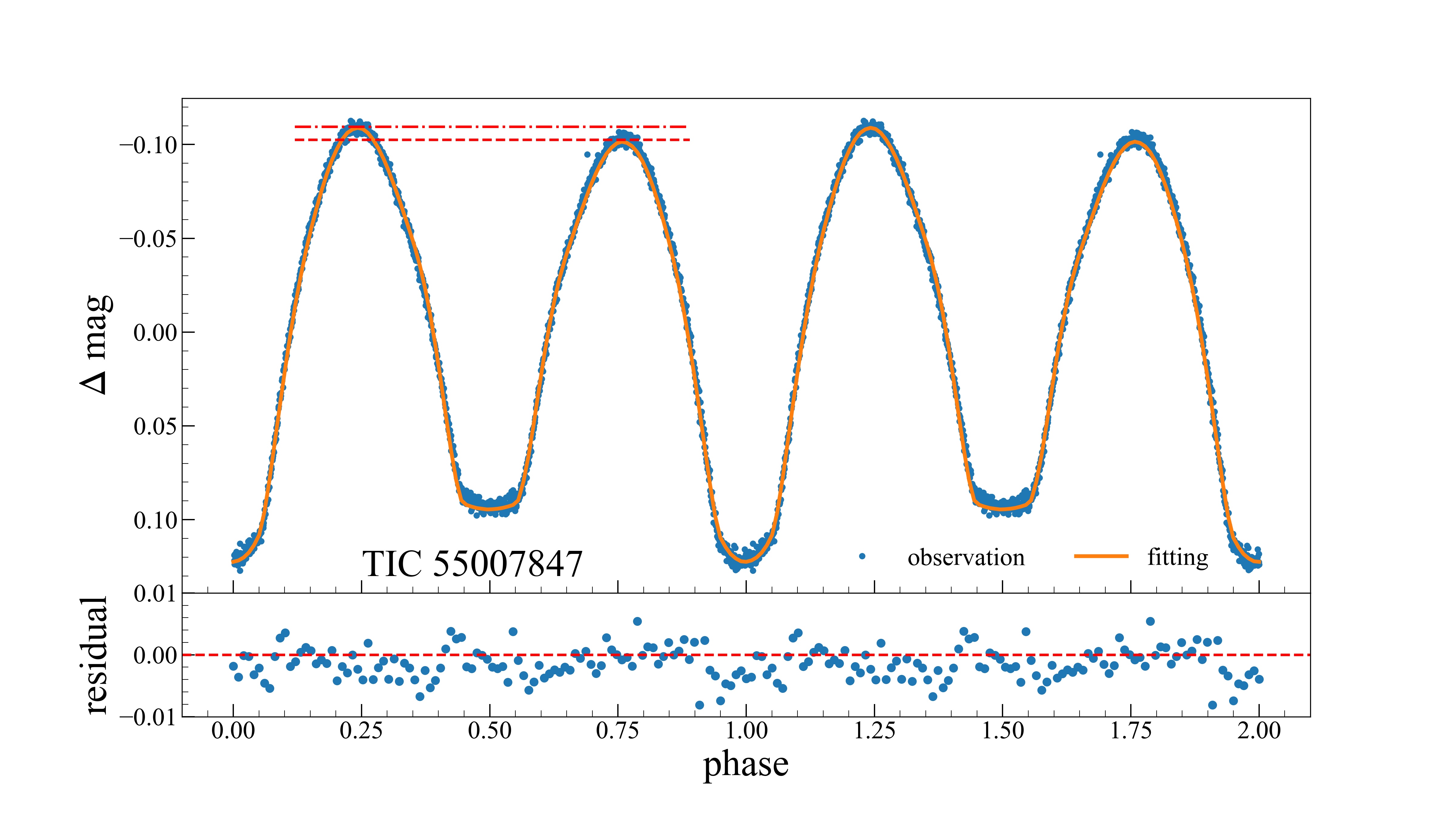}
    \caption{Light curve for the object TIC 55007847. Blue dots are the photometric data from TESS survey, and the orange solid line is the photometric solution. The two red dash line shows the O'Conell effect on the light curve.}
    \label{fig:figure 4}
\end{figure*}

Due to an asymmetric maximum of the light curve (see \textbf{Fig \ref{fig:figure 4}}), the O'Connell effect \citep{1951PRCO....2...85O} is included in our solution by adding a star spot. 
We then have some additional parameters 
including the temperature ratio of the spot to the primary star, the size, and the location ($long$ and $col$) of the spot.
The fitting result is shown in \textbf{Fig \ref{fig:figure 4}} with the solid orange line. The MCMC results are shown in \textbf{Fig \ref{fig:figure 5}},
and the corresponding parameters are presented in \textbf{Table \ref{tbl: Table 3}}.

\subsection{TIC 63597006}
\begin{figure*}
    \centering
    \includegraphics[width=0.85\textwidth]{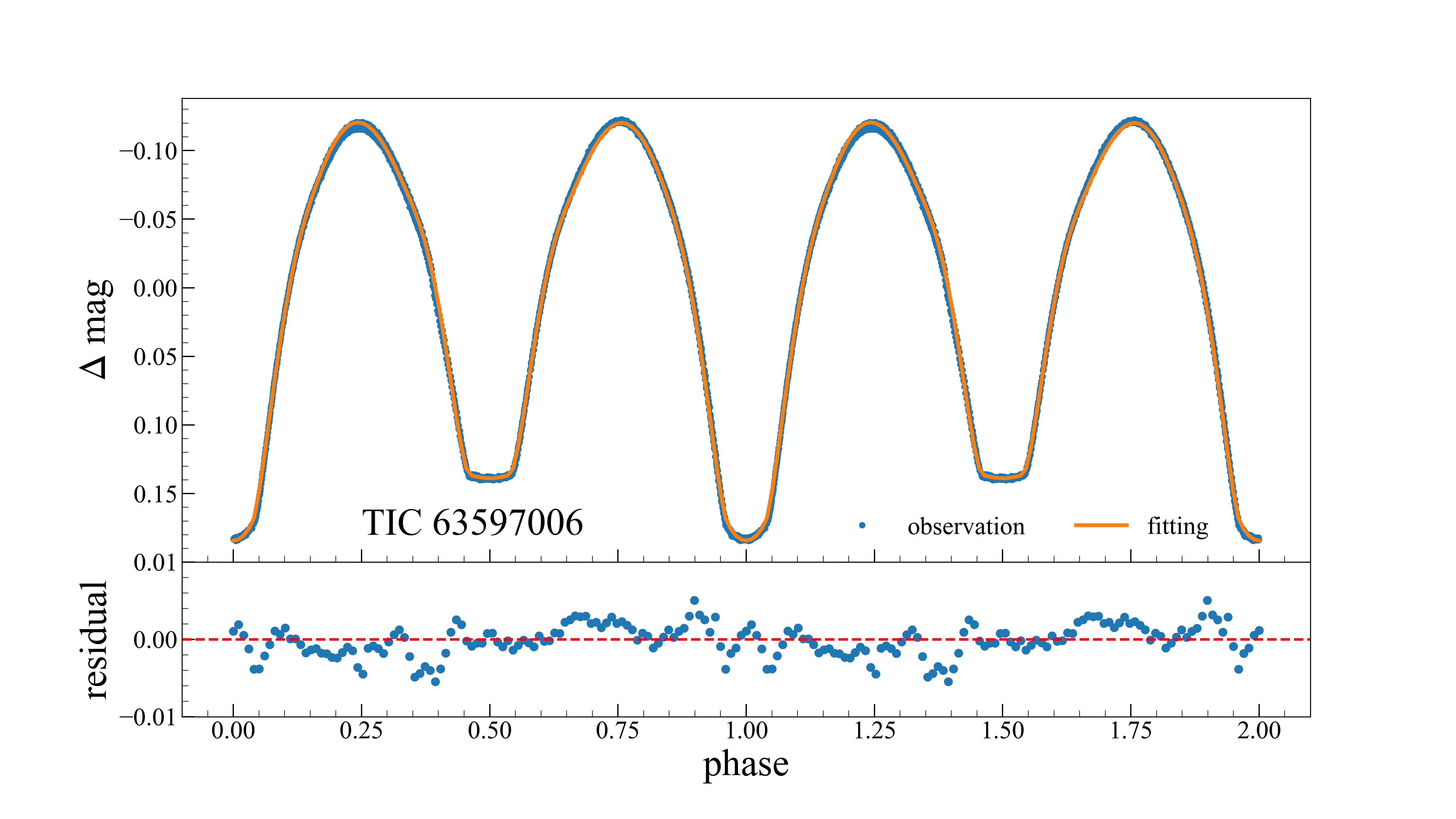}
    \caption{The photometric result of TIC 63597006. Blue dots are the photometric data from the TESS survey, and the orange solid line is the fitting result. The photometric data do not show an obvious O'Connell effect. Therefore, the model does not include the parameters of a star spot. According to the previous result, TIC 63597006 is a three-bodies system. So, we add a third light fraction to our model.}
    \label{fig:figure 6}
\end{figure*}

For this object, we set $T_1=6576\ \rm K$ from Gaia DR3 as the $T_{\rm eff}$ due to the lack of available spectra. TIC 63597006 has no obvious O'Connell effect on the light curve as shown in \textbf{Fig \ref{fig:figure 6}}. We therefore do not include star spots in the fitting. As mentioned above, however, there is a periodic signal in the O-C curve which could be caused by a third body. We then add a third body in our fitting by including the third light fraction $L_3$. The results are shown in \textbf{Fig \ref{fig:figure 6}} and \textbf{Fig \ref{fig:figure 7}}, and the details of the parameters are given in Table 3. The contribution of the third body is about $6\%$ based on the solution of \textbf{Fig \ref{fig:figure 7}}.

\begin{deluxetable*}{lclc|lclc} \label{tbl: Table 3}
\tablewidth{40pt} 
\tablecaption{The Fitting Parameters for TIC 55007847 and 
 TIC 63597006. TIC 55007847 is included a star spot in the fitting model, because the O'Connell effect is observed on the light. The O-C shows TIC 63597006 is a three-bodies system. Therefore, TIC 63597006 fitting model includes the parameters of a third light fraction.}
\tablehead{\multicolumn{4}{c}{TIC 55007847} & \multicolumn{4}{c}{TIC 63597006} \\
\hline
\colhead{Parameters} & \colhead{Value} & \colhead{Star Spots} & \colhead{Value} & \colhead{Parameters} & \colhead{Value} & \colhead{Third Light} & \colhead{Value}
} 
\startdata 
$T_{\rm p} (K)$& $7141^{+29}_{-29}$ & $T_{\rm spot}/T_{\rm p}$ & $0.85^{+0.04}_{-0.06}$ & $T_{\rm p} (K)$& $6576^{+28}_{-27}$ & $L_3$ & $0.066^{+0.01}_{-0.03}$\\
$incl (\degree)$ & $75.943^{+0.61}_{-0.62}$ & $col (\degree)$ & $93.53^{+32.64}_{-25.98}$ & $incl (\degree)$ & $79.165^{+0.39}_{-1.21}$ & {...} & {...}\\
$q$ & $0.08^{+0.002}_{-0.002}$ & $long (\degree)$ & $276.59^{+14.97}_{-21.52}$ & $q$ & $0.15^{+0.004}_{-0.009}$ & {...} & {...}\\
$f$ & $0.48^{+0.06}_{-0.04}$ & $R_{\rm spot} (\degree)$ & $23.24^{+5.84}_{-3.76}$ & $f$ & $0.15^{+0.02}_{-0.02}$ & {...} & {...}\\
$T_{\rm s}/T_{\rm p}$ & $0.99^{+0.004}_{-0.004}$ & {...} & {...} & $T_{\rm s}/T_{\rm p}$ & $0.97^{+0.002}_{-0.001}$ & {...} & {...}\\ 
$l_{\rm 2}/l_{\rm 1}$ & $0.1132$ & {...} & {...} & $l_{\rm 2}/l_{\rm 1}$ & $0.1739$ & {...} & {...}\\ 
$r_{\rm 1}/a$ & $0.6081$ & {...} & {...} & $r_{\rm 1}/a$ & $0.5503$ & {...} & {...}\\ 
$r_{\rm 2}/a$ & $0.2090$ & {...} & {...} & $r_{\rm 2}/a$ & $0.2402$ & {...} & {...}\\ 
[2pt]
\hline
\enddata
\end{deluxetable*}

\section{Estimation of the absolute Parameters} \label{sec:method}
According to the light curve solutions and the method proposed by \citet{2022MNRAS.510.5315P}, the absolute parameters of TIC 55007847 and TIC 63597006 are calculated in this section. 

\begin{figure*}
    \centering
    \includegraphics[width=1.05\textwidth]{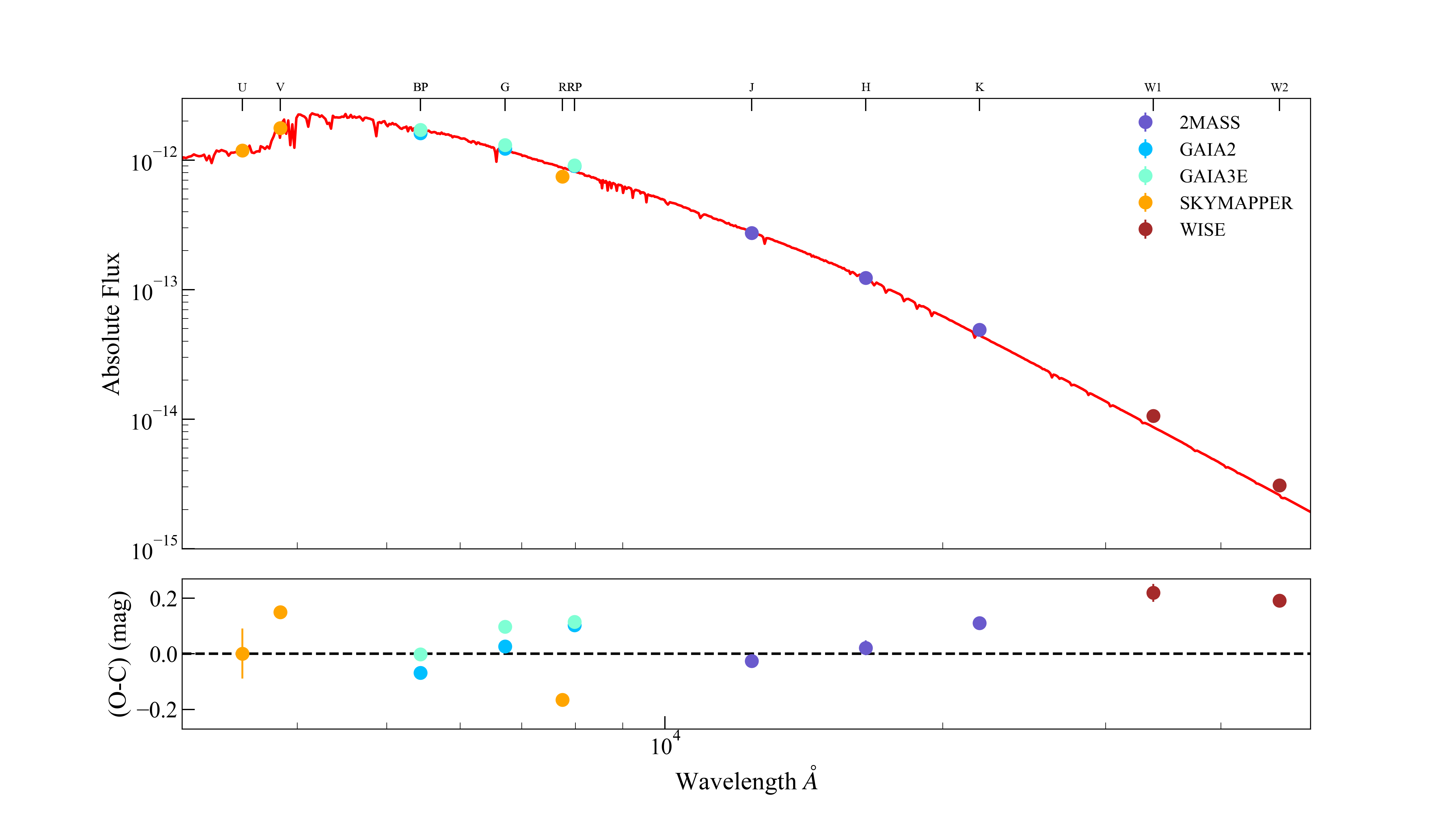}
    \caption{The SED of TIC 63597006. We use data from SkyMapper (U V R), Gaia DR2 (BP G RP), Gaia EDR3 (BP G RP), 2MASS (J H K) and WISE (W1 W2). The error bars are smaller than the symbols' sizes}
    \label{fig:figure 8}
\end{figure*}

Firstly, the absolute magnitude is derived through the parallax from Gaia DR3 by using the equation: 
\begin{equation}\label{eq:eq10}
    M_\mathrm{v} = V - 5\mathrm{log}(d) + 5 - A_\mathrm{v},
\end{equation}
where $M_{\rm v}$ is the absolute magnitude of the binary, $V$ is the V-band magnitude, $d$ is the distance, and $A_\mathrm{v}$ is the extinction.
The method of \citet{2022MNRAS.510.5315P} is based on the measurement of total luminosity. To acquire a more accurate total luminosity, the luminosity ought to be obtained from the phase that both two components face to the detector. For TIC 55007847, we obtain the total magnitude in the V-band at the maximum phase (representing the sum of luminosity for both components) of the light curve from the All-Sky Automated Survey for Supernovae (ASAS-SN) survey with the 11.35 mag. Regarding TIC 63597006, as there is no V-band light curve available from the ASAS-SN survey, we estimate its absolute parameter using the mean value of the V-band magnitude from Hipparcos. The distances are $859.54\ \rm pc$ and $205.82\ \rm pc$ for TIC 55007847 and TIC 63597006, respectively.
For TIC 55007847, $A_\mathrm{v}=0.632$ is derived from 3D dust map \citep{2018JOSS....3..695G,2018MNRAS.478..651G}. Since the location of TIC 63597006 is out of range of the 3D dust map, we cannot directly acquire the extinction. We then perform the SED (Spectral Energy Distribution) analysis to determine the extinction of TIC 63597006. In the fitting process, TIC 63597006 is treated as a single star with fixed values for distance and effective temperature. The Kurucz grid \citep{2018ASPC..515...47K} and multi-band photometry data from 2MASS, Gaia DR2, Gaia EDR3, SkyMapper, and WISE are employed for fitting. \textbf{Fig \ref{fig:figure 8}} shows the fitting results which gives $A_\mathrm{v} = 0$.

We then treat the primary star and the secondary star individually. The absolute magnitudes of each component are written as:

\begin{equation}\label{eq:eq11}
    M_\mathrm{v(1,2)} - M_\mathrm{v} = -2.5\times \mathrm{log}(\frac{l_{(1,2)}}{l})
\end{equation}
where luminosity ratio (${l_2}/{l_1}$) is obtained from the photometric solution. The total luminosity $l$ is set as 1. Therefore the luminosity ratio of primary star ${l_1}/l$ is ${1}/{(1+l_2/l_1)}$, and the luminosity ratio of secondary star is ${(l_2/l_1)}/{(1+l_2/l_1)}$. The bolometric correction (BC) is from \citet{2013ApJS..208....9P}. 

\begin{equation}\label{eq:eq12}
    M_\mathrm{bol(1,2)} = M_\mathrm{v(1,2)} + \mathrm{BC_{(1,2)}}
\end{equation}

The empirical relation of the bolometric magnitude with luminosity is written as 

\begin{equation}\label{eq:eq13}
    M_\mathrm{bol (1,2)} - M_\mathrm{bol,\odot} = -2.5\mathrm{log}(\frac{L_{(1,2)}}{L_{\odot}})
\end{equation}

The value of $M_\mathrm{bol,\odot}$ is set to 4.73 mag \citep{2010AJ....140.1158T} in \textbf{Eq (\ref{eq:eq13})}. The radii of two components can be estimated by Black body radiation.
\begin{equation}
    L_{(1,2)} = 4\pi R_{(1,2)}^2 \sigma T_{(1,2)}^4
\end{equation}

The major semi-axis $a$ can be derived from the known ($r_{(1,2)}/a$) from the photometric solutions. Finally, Kepler's Third Law (\textbf{Eq (\ref{eq:eq14})}) is used to calculate the sum of mass:
\begin{equation}\label{eq:eq14}
    \frac{a^3}{G(M_1 + M_2)} = \frac{P^2}{4\pi^2}
\end{equation}

Now we have the mass ratio $q$ from fitting models and the value of the sum of mass $M_1 + M_2$. We then can acquire the mass of each component. In addition, \citet{2022MNRAS.512.1244C} use the Main Sequence (MS) mass-luminosity relation (MLR) to estimate masses of primary star for 30 low mass ratio contact binaries. We also estimate our samples in the same method, because both two targets are located on the MS. The empirical relation is shown below:
\begin{equation}\label{eq:eq15}
    \frac{L_1}{L_\odot} = 1.49\left(\frac{M_1}{M_\odot}\right)^{4.216},
\end{equation}
where the luminosity of primary component ($L_1$) is obtained from the \textbf{Eq (\ref{eq:eq13})}, and the luminosity and mass are both in solar unit \citep{2013MNRAS.430.2029Y}.
We first obtain the mass of the primary component from \textbf{Eq (\ref{eq:eq15})}, then that of the secondary using the mass ratio from the photometric solution. 

However, when a binary system is in contact, that the energy will transfer from the more massive one to the less massive one. The luminosity of the primary star (the massive one) will be inevitably underestimated in observations and the secondary star is over-luminous. Therefore, it will also underestimate the mass of the primary star using the MS MLR. \citet{2021ApJS..254...10L} presented a MLR for contact binaries by the statistical study of 700 individually studied W UMa stars as \textbf{Eq (\ref{eq:eq16})}. We also derive the mass of the primary star by this formula.

\begin{equation} \label{eq:eq16}
    \mathrm{log} \left(\frac{L_1}{L_\odot}\right) = (2.92 \pm 0.11)\mathrm{log} \left(\frac{M_1}{M_\odot}\right) + (0.01\pm0.02)
\end{equation}

And the absolute parameters of two targets are summarized in \textbf{Table \ref{tbl: Table 4}}. The uncertainties are given by the propagation of uncertainties.

\begin{deluxetable}{lll|llBccccc} \label{tbl: Table 4}
\tablewidth{40pt} 
\tablecaption{Absolute parameters of two targets.}
\tablehead{
\multicolumn{3}{c}{TIC 55007847} & \multicolumn{2}{c}{TIC 63597006}\\
\colhead{Parameters} & \colhead{Primary} & \colhead{Secondary} & \colhead{Primary} & \colhead{Secondary} 
} 
\startdata 
$L\ (L_\odot)$ & $15.95 \pm 0.003$ & $1.81 \pm 0.014$ & $13.25 \pm 0.004$ & $2.14 \pm 0.016$ \\
$R\ (R_\odot)$ & $2.55 \pm 0.025$ & $0.86 \pm 0.009$ & $2.80 \pm 0.024$ & $1.18 \pm 0.012$ \\ 
$M\ (M_\odot)$ & $1.75 \pm 0.001$ & $0.14 \pm 0.003$ & $1.67 \pm 0.001$ & $0.23 \pm 0.003$ \\
$M\ (M_\odot)^*$ & $2.43 \pm 0.046$ & $0.20 \pm 0.015$ & $3.30 \pm 0.051$ & $0.47 \pm 0.021$ \\
$M\ (M_\odot)^{**}$ & $2.56 \pm 0.016$ & $0.20 \pm 0.005$ & $2.40 \pm 0.016$ & $0.36 \pm 0.022$ \\
$M_{\rm bol}\ ({\rm mag})$ & $1.72 \pm 0.02$ & $4.08 \pm 0.02$ & $1.92 \pm 0.03$ & $3.90 \pm 0.03$ \\
$\mathrm{log}\ g\ ({\rm cgs})$ & $3.86 \pm 2.15 $ & $3.70 \pm 2.22$ & $3.76 \pm 1.99$ & $3.66 \pm 2.03$ \\
$\mathrm{BC}$ & $-0.01 $ & $-0.01$ & $-0.04$ & $-0.05$ \\[2pt]
\hline
\enddata
\tablecomments{Mass with one star mark is derived from the solution of total mass $(M_\mathrm{(total)} = M_1 + M_2)$ and mass ratio $(q = M_2/M_1)$. Mass with two star marks is derived from the \textbf{Eq (\ref{eq:eq16})}}
\end{deluxetable}

\begin{figure*}[ht]
    \centering
    \includegraphics[width=0.75\textwidth]{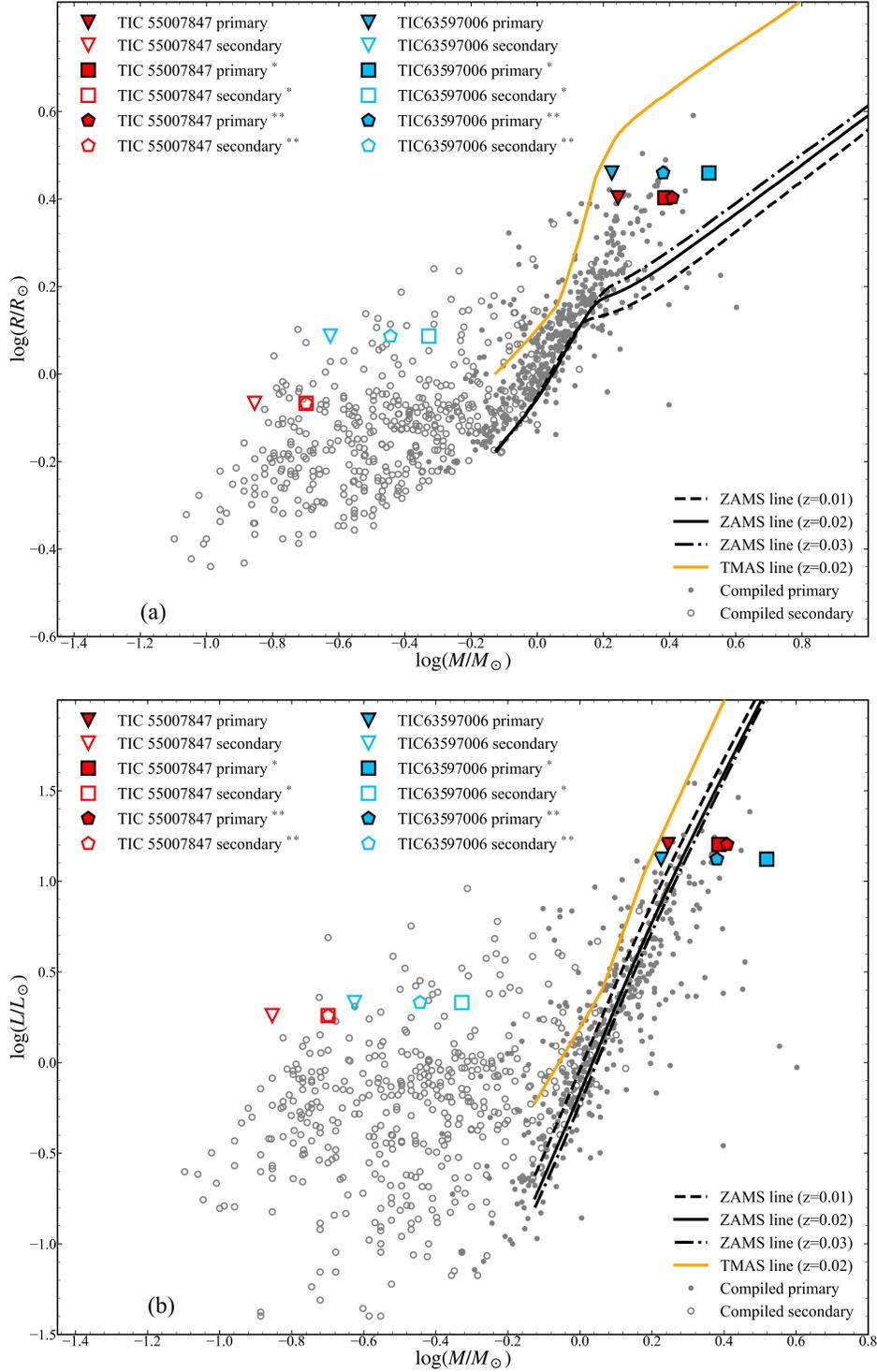}
    \caption{Comparison of our sample with low mass-ratio samples from CSS \citep{2022MNRAS.512.1244C} and individually studied W UMa samples \citep{2021ApJS..254...10L} on the mass-radius (panel (a)) and mass-luminosity (panel (b)) diagrams. The black lines are the Zero Age Main Sequence with different metallicities ($Z$). The dashed line, solid line, and dashed dotted line represent the ZAMS with $Z$=0.01, $Z$=0.02, and $Z$=0.03, respectively. The orange line represents the Terminal Age Main Sequence (TAMS). The grey solid and open circles are the primary stars and secondary stars of the compiled samples (W Uma catalog \citep{2021ApJS..254...10L} and the 30 new low mass ratio contact binaries from CSS \citep{2021ApJS..254...10L}). The red symbols represent the TIC 55007847 and the blue symbols represent TIC 63597006. The solid symbols and the open symbols represent the primary stars and the secondary stars, respectively. The different symbols represent the masses obtained from different methods, i.e the triangles, the rectangles, and pentagons are the mass from \textbf{Eq (\ref{eq:eq15})}, the method of \citet{2022MNRAS.510.5315P}, and the \textbf{Eq (\ref{eq:eq16})}.}
    \label{fig:figure 10}
\end{figure*}

\textbf{Fig \ref{fig:figure 10}} shows the mass-radius (panel (a)) and mass-luminosity (panel (b)) relations for both of TIC 55007847 (red symbols) and TIC 63597006 (blue symbols). The solid symbols and open symbols represent the primary components and secondary components, respectively. The different symbols represent the masses from different methods, i.e the triangles, the rectangles and the pentagons are the masses from \textbf{Eq (\ref{eq:eq15})}, the method of \citet{2022MNRAS.510.5315P} and \textbf{Eq (\ref{eq:eq16})}, respectively. If we assume that the primary masses derived by the MLR from contact binary systems are more accurate than those estimated by the MLR of single MS stars, the primary star masses of  TIC~55007847 and TIC~63597006 would be underestimated by $0.81\ M_\odot$ and $0.76\ M_\odot$, respectively. However, the estimated primary masses of these two objects given by the MLR of single MS stars can still fall into the MLR distribution of the sample of 700 contact binaries. This seems to indicate that the estimates from the MLR of single MS stars are still acceptable. In this work, we adopt the MLR given by contact binaries of \citet{2021ApJS..254...10L} for the following study.

\section{The stability analysis} \label{sec:dis}

As mentioned in Section 1, contact binaries with extremely low mass ratios are likely progenitors of some merging stars such as FK Com stars and BSs. According to \citet{1980A&A....92..167H}, the \textbf{tidal} instability (Darwin instability) occurs when the total spin angular momentum ($J_{\rm s}$) is larger than $1/3$ of the orbital angular momentum ($J_{\rm o}$), saying, the binary will merge into a single star when with $J_{\rm s}/J_{\rm o}>1/3$. 
We investigate the evolutionary statuses for the two targets here.

\begin{figure*}[t]
  \centering
    \subfigure{
   \includegraphics[scale=0.25]{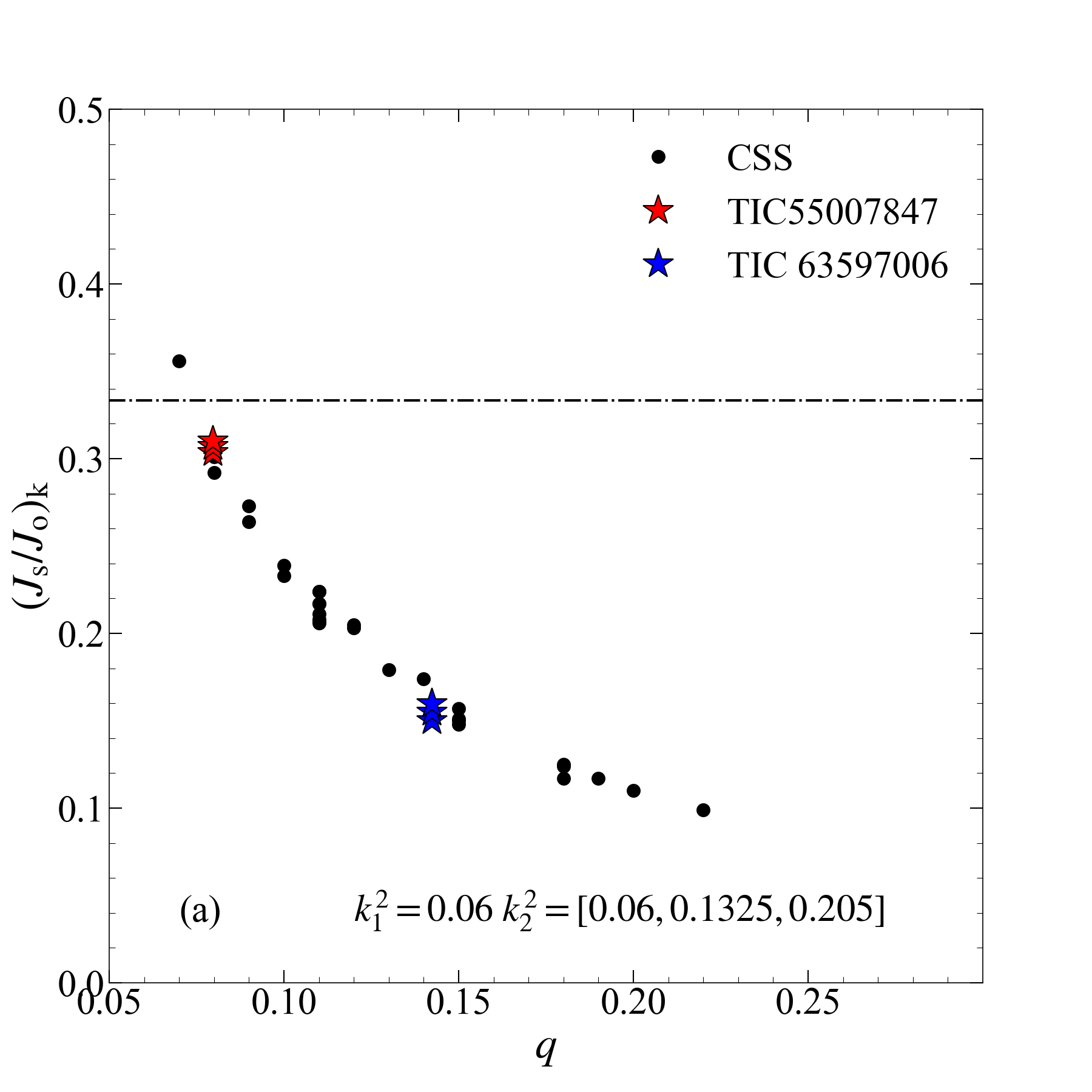}
  }
  \subfigure{
   \includegraphics[scale=0.25]{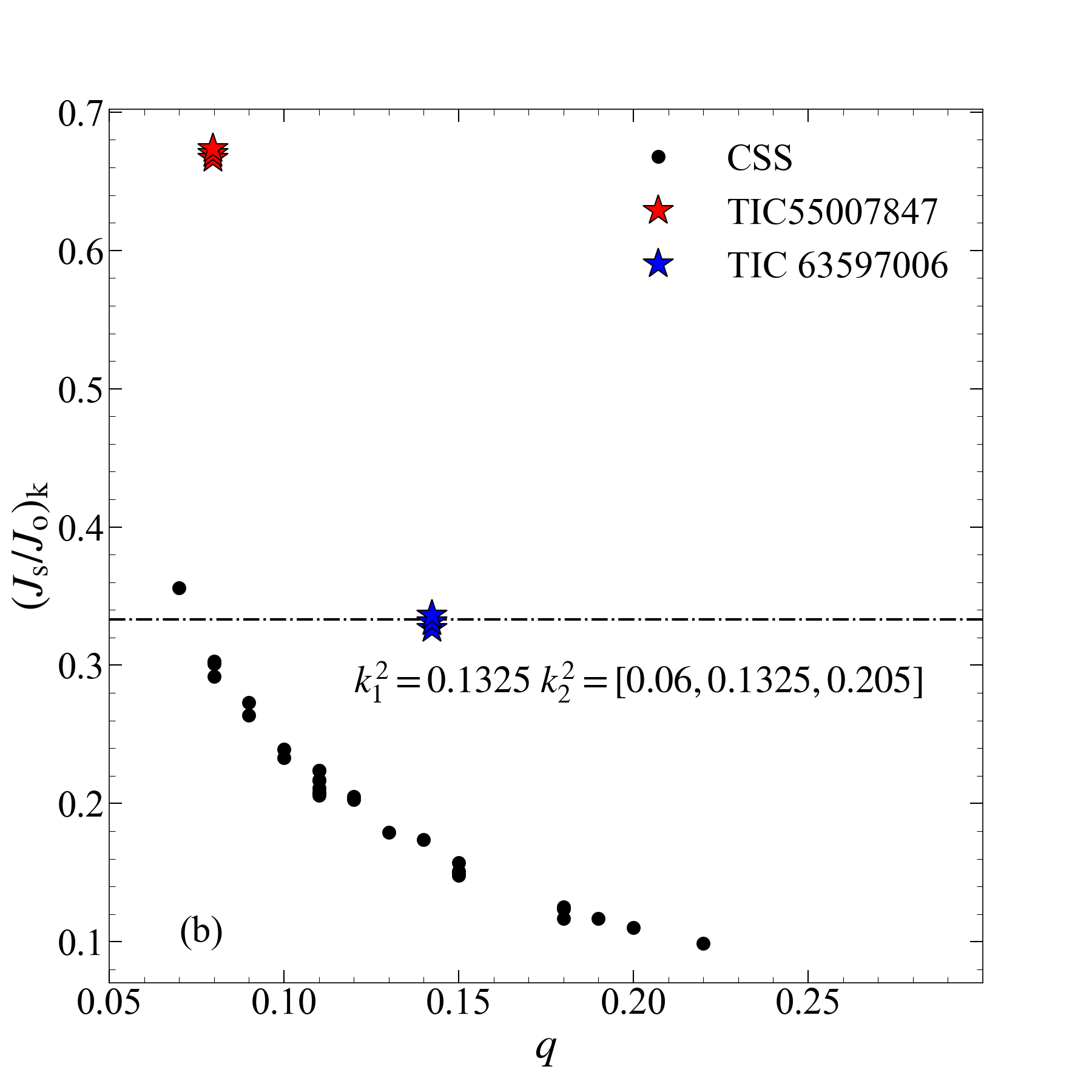}
  }
    \subfigure{
   \includegraphics[scale=0.25]{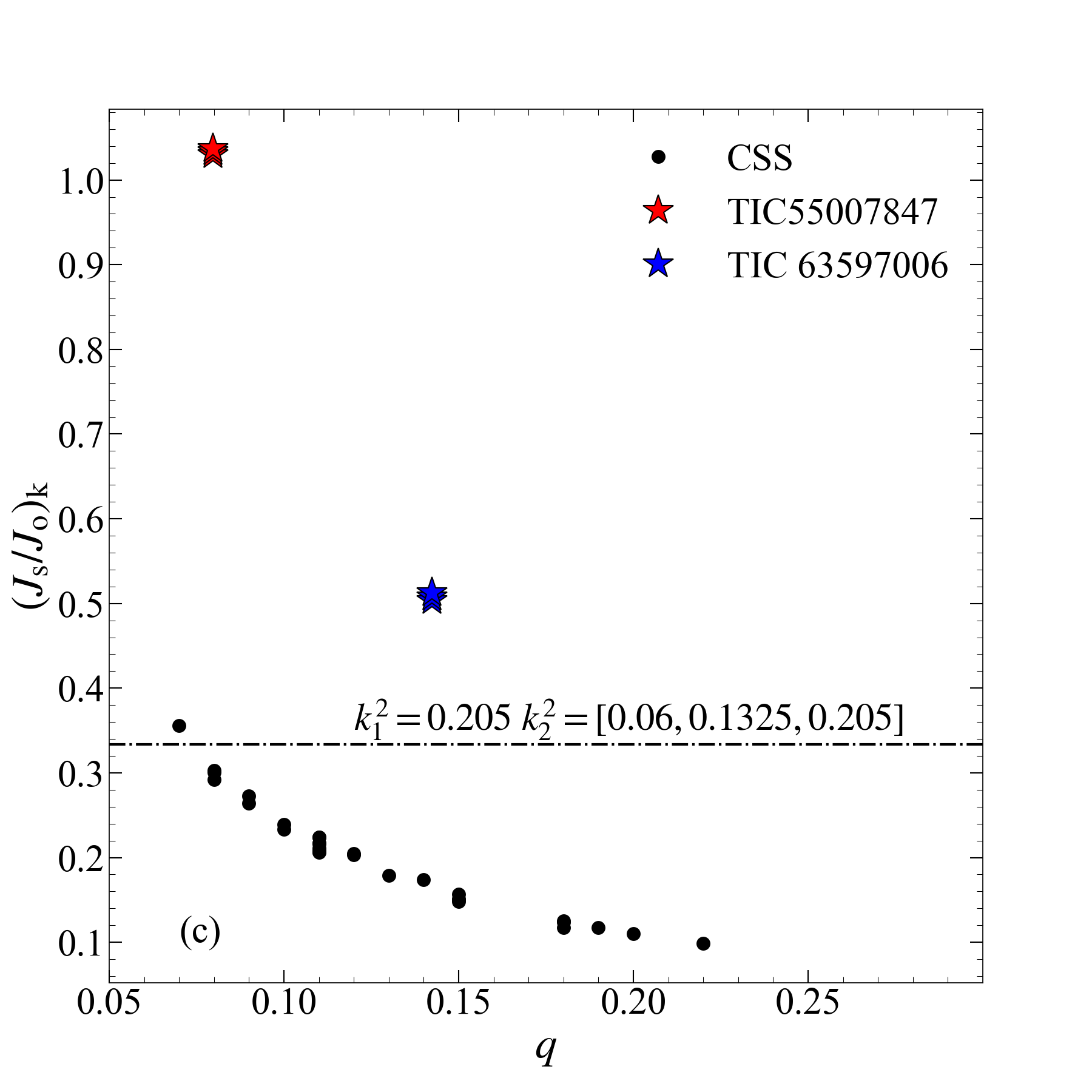}
  }
   \subfigure{
   \includegraphics[scale=0.25]{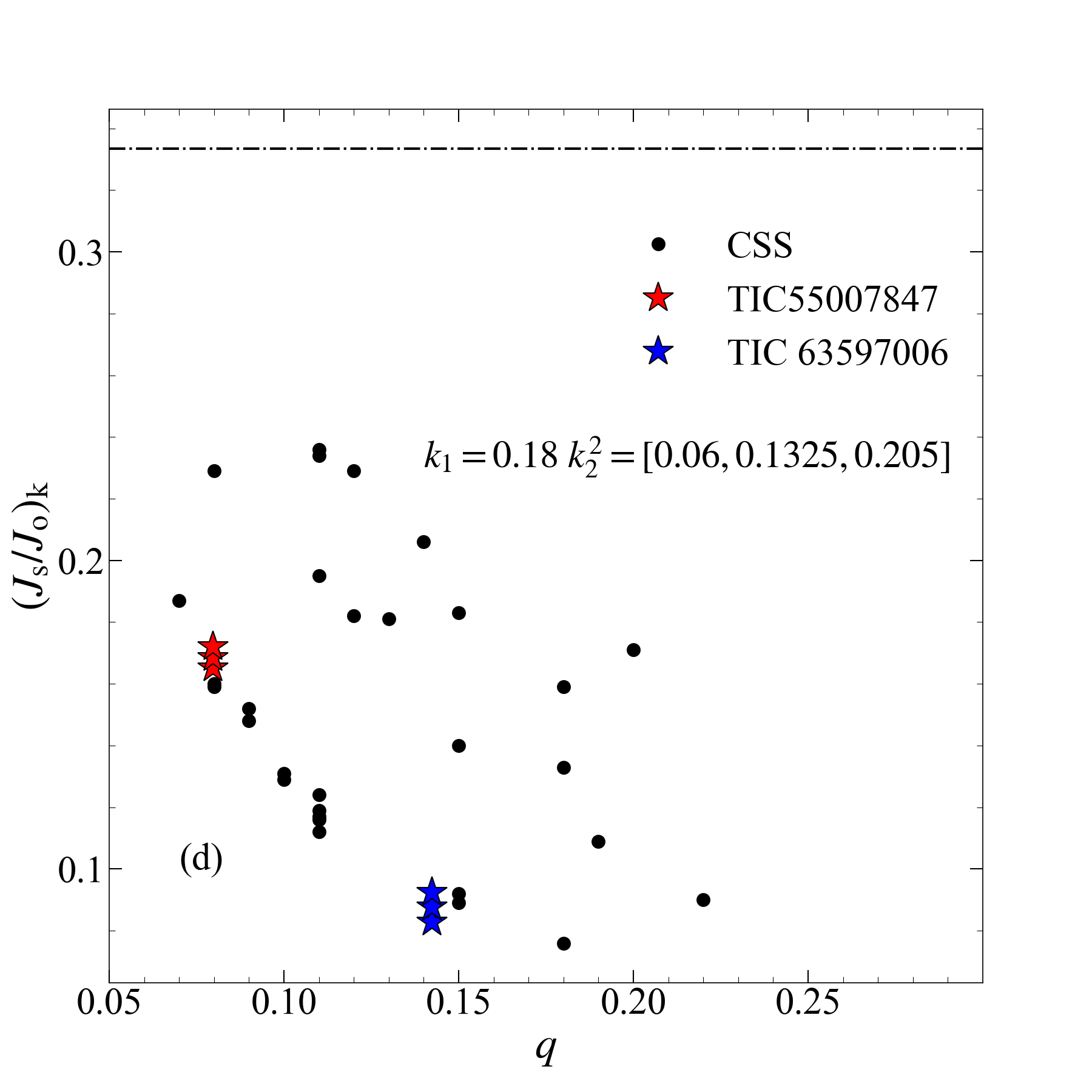}
  }
  \centering
  \caption{
  This figure shows the merger status of TIC 55007847 and TIC 63597006 when we use the different dimensionless gyration radii. According to \citet{2022MNRAS.510.5315P}, the $k^2$ ranges from 0.06 to 0.205. We set $k_1^2 = 0.06$ (fully radiative) in panel (a). The $k_1^2$ is 0.1325 (a mid status) in panel (b), and the $k_1^2$ is 0.205 (fully convective) in panel (c). In each panel, the bottom stars are the status of $k_2^2=0.06$ and the top stars are the $k_2^2=0.205$. The red stars are the TIC 55007847 and the blue stars are the TIC 63597006. Moreover, when $M > 1.4$ solar mass, $k_1$ can be regarded as a constant value of 0.18, for our targets. The panel (d) shows the different merger statuses for $k_2^2 = 0.06$, $k_2^2 = 0.1325$ and $k_2^2 = 0.205$.}\label{fig:figure 12}
\end{figure*}

Generally, the value of $J_{\rm s}/J_{\rm o}$ can be calculated by \citep{2006MNRAS.369.2001L}: 

\begin{equation}\label{eq:eq17}
    \frac{J_{\rm s}}{J_{\rm o}} = \frac{(1+q)}{q} (k_1 r_1)^2 \left[ 1+q(\frac{k_2}{k_1})^2(\frac{r_2}{r_1})^2 \right],
\end{equation}
where $k_1$, and $k_2$ are the dimensionless gyration radii of the primary star and secondary star. Due to $k_1$ and $k_2$ being unknown in the equation, the estimation of the value of gyration radii will influence the result directly. According to the \citet{2022MNRAS.512.1244C}, the value of gyration radii distributes can be 0.06 (sun-like) \citep{1995ApJ...444L..41R,2006MNRAS.369.2001L}, 0.075 (fully radiative) to 0.205 (fully convective). Therefore, we first assume the $k_1^2 = 0.06$ (sun-like), and $k_2^2$ ranges from 0.06 to 0.205, the results are shown in panel (a) of \textbf{Fig \ref{fig:figure 12}}, in this panel the black dots are the sample from \citet{2022MNRAS.512.1244C}, their $J_{\rm s}/J_{\rm o}$ are derived by assuming $k_1^2 = k_2^2 = 0.06$ (sun-like), and the blue and red star stars are TIC 55007847 and TIC 63597006 respectively, while the bottom and top red and blue stars correspond to the results with $k_2^2=0.06$ and $k_2^2=0.205$, respectively. It can be seen that when assuming $k_1^2 = 0.06$, with $k_2^2$ varying from 0.06 to 0.205, TIC 55007847 is located close to the limit, while TIC 63597006 remains within the constraint. When we assume that $k_1^2 = 0.1325$ (a mid-status between the fully radiative and the fully convective) and $k_2^2$ ranges from 0.06 to 0.205 TIC 55007847 is over the merger threshold of 1/3 in panel (b). When $k_1^2 = 0.205$ (fully convective) and $k_2^2$ ranges from 0.06 to 0.205, both TIC 55007847 and TIC 63597006 will merge in panel (c). Moreover, when $M > 1.4\ M_\odot$, $k_1$ can be regarded as a constant value of 0.18 \citep{2022MNRAS.512.1244C}, for our targets, the masses of primary components are all greater than $1.4 M_\odot$; hence, we also set $k_1$ = 0.18 with $k_2^2$ varying from 0.06 to 0.205 to analyze the merger state. The results are shown in panel (d). the black dots are the sample from \citet{2022MNRAS.512.1244C}, their $J_{\rm s}/J_{\rm o}$ are derived by assuming $k_1 = 0.18$, and the blue and red stars are TIC 55007847 and TIC 63597006 respectively, while the bottom and top red and blue markers correspond to $k_2^2=0.06$ and $k_2^2=0.205$. Neither TIC 55007847 nor TIC 63597006 can merge into a fast-rotating single star.

\section{Conclusion} \label{sec:con}

In this paper, we make a light curve analysis of two contact binaries with low mass ratios identified from TESS. 
We obtain the precise orbital periods for the two objects by using the O-C method, and find an obvious periodic signal in the O-C diagram for the object of TIC 63597006. We suggest such a periodic signal is from the third light, and the contribution of the third light is about 6$\%$.

We further obtain the relative parameters from the light curve analysis 
by using the MCMC with PHOEBE method and derive the absolute parameters of the two targets. The study shows the mass ratio of 0.08 and 0.14 for 
TIC 55007847 and TIC 63597006, respectively. 

Using the masses derived from the empirical luminosity-mass relation, 
we investigate the evolutionary status of the objects. 
We find that, in the case of $k_1^2 = k_2^2 = 0.06$,  
the value of $J_{\rm s}/J_{\rm o}$ is very close to the Darwin instability line ($J_{\rm s}/J_{\rm o}=1/3$) for TIC 55007847. 
It means that this object is likely to enter into the merge process soon. 
The object TIC 63597006 is relatively stable at present.

\section{Acknowledgments} \label{acknowledgments}

This work is supported by the National Natural Science Foundation of China (grants Nos. 12125303, 12288102, and 12090040/3), the National Key R$\&$D Program of China (grant No.2021YFA1600403), the Yunnan Fundamental Research Projects (grants Nos. 202201BCO70003), the International Centre of Supernovae, Yunnan Key Laboratory (No.202302AN360001) and the Yunnan Revitalization Talent Support Program -- Science $\&$ Technology Champion Project (grant NO.202305AB350003). This work is also supported by the China Manned Space Project of No.CMS-CSST-2021-A10. This work is partly supported by the Chinese Natural Science Foundation (Nos. 12103088).

\bibliographystyle{aasjournal} 
\bibliography{FP}

\begin{thebibliography}{}
\expandafter\ifx\csname natexlab\endcsname\relax\def\natexlab#1{#1}\fi
\providecommand{\url}[1]{\href{#1}{#1}}
\providecommand{\dodoi}[1]{doi:~\href{http://doi.org/#1}{\nolinkurl{#1}}}
\providecommand{\doeprint}[1]{\href{http://ascl.net/#1}{\nolinkurl{http://ascl.net/#1}}}
\providecommand{\doarXiv}[1]{\href{https://arxiv.org/abs/#1}{\nolinkurl{https://arxiv.org/abs/#1}}}

\bibitem[{{Applegate}(1992)}]{1992ApJ...385..621A}
{Applegate}, J.~H. 1992, \apj, 385, 621, \dodoi{10.1086/170967}

\bibitem[{{Binnendijk}(1970)}]{1970VA.....12..217B}
{Binnendijk}, L. 1970, Vistas in Astronomy, 12, 217,
  \dodoi{10.1016/0083-6656(70)90041-3}

\bibitem[{{Binnendijk}(1977)}]{1977VA.....21..359B}
---. 1977, Vistas in Astronomy, 21, 359, \dodoi{10.1016/0083-6656(77)90022-8}

\bibitem[{{Christopoulou} {et~al.}(2022){Christopoulou}, {Lalounta},
  {Papageorgiou}, {Ferreira Lopes}, {Catelan}, \&
  {Drake}}]{2022MNRAS.512.1244C}
{Christopoulou}, P.-E., {Lalounta}, E., {Papageorgiou}, A., {et~al.} 2022,
  \mnras, 512, 1244, \dodoi{10.1093/mnras/stac534}

\bibitem[{{Ding} {et~al.}(2021){Ding}, {Ji}, \& {Li}}]{2021PASJ...73..786D}
{Ding}, X., {Ji}, K.-F., \& {Li}, X.-Z. 2021, \pasj, 73, 786,
  \dodoi{10.1093/pasj/psab042}

\bibitem[{{Foreman-Mackey} {et~al.}(2013){Foreman-Mackey}, {Hogg}, {Lang}, \&
  {Goodman}}]{2013PASP..125..306F}
{Foreman-Mackey}, D., {Hogg}, D.~W., {Lang}, D., \& {Goodman}, J. 2013, \pasp,
  125, 306, \dodoi{10.1086/670067}

\bibitem[{{Green}(2018)}]{2018JOSS....3..695G}
{Green}, G.~M. 2018, The Journal of Open Source Software, 3, 695,
  \dodoi{10.21105/joss.00695}

\bibitem[{{Green} {et~al.}(2018){Green}, {Schlafly}, {Finkbeiner}, {Rix},
  {Martin}, {Burgett}, {Draper}, {Flewelling}, {Hodapp}, {Kaiser}, {Kudritzki},
  {Magnier}, {Metcalfe}, {Tonry}, {Wainscoat}, \&
  {Waters}}]{2018MNRAS.478..651G}
{Green}, G.~M., {Schlafly}, E.~F., {Finkbeiner}, D., {et~al.} 2018, \mnras,
  478, 651, \dodoi{10.1093/mnras/sty1008}

\bibitem[{{Huang}(1966)}]{1966ARA&A...4...35H}
{Huang}, S.-S. 1966, \araa, 4, 35, \dodoi{10.1146/annurev.aa.04.090166.000343}

\bibitem[{{Hut}(1980)}]{1980A&A....92..167H}
{Hut}, P. 1980, \aap, 92, 167

\bibitem[{{IJspeert} {et~al.}(2021){IJspeert}, {Tkachenko}, {Johnston},
  {Garcia}, {De Ridder}, {Van Reeth}, \& {Aerts}}]{2021A&A...652A.120I}
{IJspeert}, L.~W., {Tkachenko}, A., {Johnston}, C., {et~al.} 2021, \aap, 652,
  A120, \dodoi{10.1051/0004-6361/202141489}

\bibitem[{{Irwin}(1952)}]{1952ApJ...116..211I}
{Irwin}, J.~B. 1952, \apj, 116, 211, \dodoi{10.1086/145604}

\bibitem[{{Koci{\'a}n}(2019)}]{2019OEJV..197.....K}
{Koci{\'a}n}, R. 2019, Open European Journal on Variable Stars, 197

\bibitem[{{Kopal}(1959)}]{1959cbs..book.....K}
{Kopal}, Z. 1959, {Close binary systems}

\bibitem[{{Kurucz}(2018)}]{2018ASPC..515...47K}
{Kurucz}, R.~L. 2018, in Astronomical Society of the Pacific Conference Series,
  Vol. 515, Workshop on Astrophysical Opacities, 47

\bibitem[{{Latkovi{\'c}} {et~al.}(2021){Latkovi{\'c}}, {{\v{C}}eki}, \&
  {Lazarevi{\'c}}}]{2021ApJS..254...10L}
{Latkovi{\'c}}, O., {{\v{C}}eki}, A., \& {Lazarevi{\'c}}, S. 2021, \apjs, 254,
  10, \dodoi{10.3847/1538-4365/abeb23}

\bibitem[{{Li} {et~al.}(2022){Li}, {Gao}, {Liu}, {Gao}, {Li}, {Chen}, \&
  {Sun}}]{2022yCat..51640202L}
{Li}, K., {Gao}, X., {Liu}, X.~Y., {et~al.} 2022, VizieR Online Data Catalog,
  J/AJ/164/202, \dodoi{10.26093/cds/vizier.51640202}

\bibitem[{{Li} \& {Zhang}(2006)}]{2006MNRAS.369.2001L}
{Li}, L., \& {Zhang}, F. 2006, \mnras, 369, 2001,
  \dodoi{10.1111/j.1365-2966.2006.10462.x}

\bibitem[{{Lindegren} {et~al.}(2018){Lindegren}, {Hern{\'a}ndez}, {Bombrun},
  {Klioner}, {Bastian}, {Ramos-Lerate}, {de Torres}, {Steidelm{\"u}ller},
  {Stephenson}, {Hobbs}, {Lammers}, {Biermann}, {Geyer}, {Hilger}, {Michalik},
  {Stampa}, {McMillan}, {Casta{\~n}eda}, {Clotet}, {Comoretto}, {Davidson},
  {Fabricius}, {Gracia}, {Hambly}, {Hutton}, {Mora}, {Portell}, {van Leeuwen},
  {Abbas}, {Abreu}, {Altmann}, {Andrei}, {Anglada}, {Balaguer-N{\'u}{\~n}ez},
  {Barache}, {Becciani}, {Bertone}, {Bianchi}, {Bouquillon}, {Bourda},
  {Br{\"u}semeister}, {Bucciarelli}, {Busonero}, {Buzzi}, {Cancelliere},
  {Carlucci}, {Charlot}, {Cheek}, {Crosta}, {Crowley}, {de Bruijne}, {de
  Felice}, {Drimmel}, {Esquej}, {Fienga}, {Fraile}, {Gai}, {Garralda},
  {Gonz{\'a}lez-Vidal}, {Guerra}, {Hauser}, {Hofmann}, {Holl}, {Jordan},
  {Lattanzi}, {Lenhardt}, {Liao}, {Licata}, {Lister}, {L{\"o}ffler},
  {Marchant}, {Martin-Fleitas}, {Messineo}, {Mignard}, {Morbidelli}, {Poggio},
  {Riva}, {Rowell}, {Salguero}, {Sarasso}, {Sciacca}, {Siddiqui}, {Smart},
  {Spagna}, {Steele}, {Taris}, {Torra}, {van Elteren}, {van Reeven}, \&
  {Vecchiato}}]{2018A&A...616A...2L}
{Lindegren}, L., {Hern{\'a}ndez}, J., {Bombrun}, A., {et~al.} 2018, \aap, 616,
  A2, \dodoi{10.1051/0004-6361/201832727}

\bibitem[{{Lu} {et~al.}(2020){Lu}, {Zhang}, {Michel}, \&
  {Han}}]{2020ApJ...901..169L}
{Lu}, H.-p., {Zhang}, L.-y., {Michel}, R., \& {Han}, X.~L. 2020, \apj, 901,
  169, \dodoi{10.3847/1538-4357/abb19b}

\bibitem[{{Lucy}(1967)}]{1967ZA.....65...89L}
{Lucy}, L.~B. 1967, \zap, 65, 89

\bibitem[{{Molnar} {et~al.}(2017){Molnar}, {Van Noord}, {Kinemuchi},
  {Smolinski}, {Alexander}, {Cook}, {Jang}, {Kobulnicky}, {Spedden}, \&
  {Steenwyk}}]{2017ApJ...840....1M}
{Molnar}, L.~A., {Van Noord}, D.~M., {Kinemuchi}, K., {et~al.} 2017, \apj, 840,
  1, \dodoi{10.3847/1538-4357/aa6ba7}

\bibitem[{{O'Connell}(1951)}]{1951PRCO....2...85O}
{O'Connell}, D.~J.~K. 1951, Publications of the Riverview College Observatory,
  2, 85

\bibitem[{{Oosterhoff}(1950)}]{1950BAN....11..217O}
{Oosterhoff}, P.~T. 1950, \bain, 11, 217

\bibitem[{{Pecaut} \& {Mamajek}(2013)}]{2013ApJS..208....9P}
{Pecaut}, M.~J., \& {Mamajek}, E.~E. 2013, \apjs, 208, 9,
  \dodoi{10.1088/0067-0049/208/1/9}

\bibitem[{{Pe{\v{s}}ta} \& {Pejcha}(2023)}]{2023A&A...672A.176P}
{Pe{\v{s}}ta}, M., \& {Pejcha}, O. 2023, \aap, 672, A176,
  \dodoi{10.1051/0004-6361/202245613}

\bibitem[{{Poro} {et~al.}(2022){Poro}, {Sarabi}, {Zamanpour}, {Fotouhi},
  {Davoudi}, {Khakpash}, {Salehian}, {Madayen}, {Foroutanfar}, {Bakhshi},
  {Mahdavi}, {Alicavus}, {Farahani}, {Sabbaghian}, {Hosseini}, {Aryaeefar}, \&
  {Hemati}}]{2022MNRAS.510.5315P}
{Poro}, A., {Sarabi}, S., {Zamanpour}, S., {et~al.} 2022, \mnras, 510, 5315,
  \dodoi{10.1093/mnras/stab3775}

\bibitem[{{Pribulla} \& {Rucinski}(2006)}]{2006AJ....131.2986P}
{Pribulla}, T., \& {Rucinski}, S.~M. 2006, \aj, 131, 2986,
  \dodoi{10.1086/503871}

\bibitem[{{Pr{\v{s}}a} {et~al.}(2016){Pr{\v{s}}a}, {Conroy}, {Horvat}, {Pablo},
  {Kochoska}, {Bloemen}, {Giammarco}, {Hambleton}, \&
  {Degroote}}]{2016ApJS..227...29P}
{Pr{\v{s}}a}, A., {Conroy}, K.~E., {Horvat}, M., {et~al.} 2016, \apjs, 227, 29,
  \dodoi{10.3847/1538-4365/227/2/29}

\bibitem[{{Pr{\v{s}}a} {et~al.}(2022){Pr{\v{s}}a}, {Kochoska}, {Conroy},
  {Eisner}, {Hey}, {IJspeert}, {Kruse}, {Fleming}, {Johnston}, {Kristiansen},
  {LaCourse}, {Mortensen}, {Pepper}, {Stassun}, {Torres}, {Abdul-Masih},
  {Chakraborty}, {Gagliano}, {Guo}, {Hambleton}, {Hong}, {Jacobs}, {Jones},
  {Kostov}, {Lee}, {Omohundro}, {Orosz}, {Page}, {Powell}, {Rappaport}, {Reed},
  {Schnittman}, {Schwengeler}, {Shporer}, {Terentev}, {Vanderburg}, {Welsh},
  {Caldwell}, {Doty}, {Jenkins}, {Latham}, {Ricker}, {Seager}, {Schlieder},
  {Shiao}, {Vanderspek}, \& {Winn}}]{2022ApJS..258...16P}
{Pr{\v{s}}a}, A., {Kochoska}, A., {Conroy}, K.~E., {et~al.} 2022, \apjs, 258,
  16, \dodoi{10.3847/1538-4365/ac324a}

\bibitem[{{Rappaport}(2011)}]{2011cxo..prop.3317R}
{Rappaport}, S. 2011, {V1309 Sco: The First Directly Observed Merging Binary},
  Chandra proposal ID 13200318

\bibitem[{{Rasio}(1995)}]{1995ApJ...444L..41R}
{Rasio}, F.~A. 1995, \apjl, 444, L41, \dodoi{10.1086/187855}

\bibitem[{{Ricker} {et~al.}(2015){Ricker}, {Winn}, {Vanderspek}, {Latham},
  {Bakos}, {Bean}, {Berta-Thompson}, {Brown}, {Buchhave}, {Butler}, {Butler},
  {Chaplin}, {Charbonneau}, {Christensen-Dalsgaard}, {Clampin}, {Deming},
  {Doty}, {De Lee}, {Dressing}, {Dunham}, {Endl}, {Fressin}, {Ge}, {Henning},
  {Holman}, {Howard}, {Ida}, {Jenkins}, {Jernigan}, {Johnson}, {Kaltenegger},
  {Kawai}, {Kjeldsen}, {Laughlin}, {Levine}, {Lin}, {Lissauer}, {MacQueen},
  {Marcy}, {McCullough}, {Morton}, {Narita}, {Paegert}, {Palle}, {Pepe},
  {Pepper}, {Quirrenbach}, {Rinehart}, {Sasselov}, {Sato}, {Seager},
  {Sozzetti}, {Stassun}, {Sullivan}, {Szentgyorgyi}, {Torres}, {Udry}, \&
  {Villasenor}}]{2015JATIS...1a4003R}
{Ricker}, G.~R., {Winn}, J.~N., {Vanderspek}, R., {et~al.} 2015, Journal of
  Astronomical Telescopes, Instruments, and Systems, 1, 014003,
  \dodoi{10.1117/1.JATIS.1.1.014003}

\bibitem[{{Ruci{\'n}ski}(1969)}]{1969AcA....19..245R}
{Ruci{\'n}ski}, S.~M. 1969, \actaa, 19, 245

\bibitem[{{Tang} {et~al.}(2021){Tang}, {Gai}, {Li}, {Yang}, \&
  {Dong}}]{2021AcASn..62...39T}
{Tang}, Y.~K., {Gai}, N., {Li}, Z.~K., {Yang}, H.~L., \& {Dong}, W.~H. 2021,
  Acta Astronomica Sinica, 62, 39

\bibitem[{{Torres}(2010)}]{2010AJ....140.1158T}
{Torres}, G. 2010, \aj, 140, 1158, \dodoi{10.1088/0004-6256/140/5/1158}

\bibitem[{{Tylenda} {et~al.}(2011){Tylenda}, {Hajduk}, {Kami{\'n}ski},
  {Udalski}, {Soszy{\'n}ski}, {Szyma{\'n}ski}, {Kubiak}, {Pietrzy{\'n}ski},
  {Poleski}, {Wyrzykowski}, \& {Ulaczyk}}]{2011A&A...528A.114T}
{Tylenda}, R., {Hajduk}, M., {Kami{\'n}ski}, T., {et~al.} 2011, \aap, 528,
  A114, \dodoi{10.1051/0004-6361/201016221}

\bibitem[{{Wadhwa} {et~al.}(2021){Wadhwa}, {Tothill}, {DeHorta}, \&
  {Filipovi{\'c}}}]{2021RAA....21..235W}
{Wadhwa}, S.~S., {Tothill}, N. F.~H., {DeHorta}, A.~Y., \& {Filipovi{\'c}}, M.
  2021, Research in Astronomy and Astrophysics, 21, 235,
  \dodoi{10.1088/1674-4527/21/9/235}

\bibitem[{{Webbink}(1976)}]{1976ApJ...209..829W}
{Webbink}, R.~F. 1976, \apj, 209, 829, \dodoi{10.1086/154781}

\bibitem[{{Yang} \& {Qian}(2015)}]{2015AJ....150...69Y}
{Yang}, Y.-G., \& {Qian}, S.-B. 2015, \aj, 150, 69,
  \dodoi{10.1088/0004-6256/150/3/69}

\bibitem[{{Yildiz} \& {Do{\u{g}}an}(2013)}]{2013MNRAS.430.2029Y}
{Yildiz}, M., \& {Do{\u{g}}an}, T. 2013, \mnras, 430, 2029,
  \dodoi{10.1093/mnras/stt028}

\bibitem[{{Zechmeister} \& {K{\"u}rster}(2009)}]{2009A&A...496..577Z}
{Zechmeister}, M., \& {K{\"u}rster}, M. 2009, \aap, 496, 577,
  \dodoi{10.1051/0004-6361:200811296}

\end{thebibliography}
\begin{appendices}
\section*{Appendix}

In this section, we show the corner diagrams of two targets. The fitting parameters are summarized in \textbf{Table \ref{tbl: Table 3}}.

\begin{figure*}[h]
    \centering
    \includegraphics[width=0.85\textwidth]{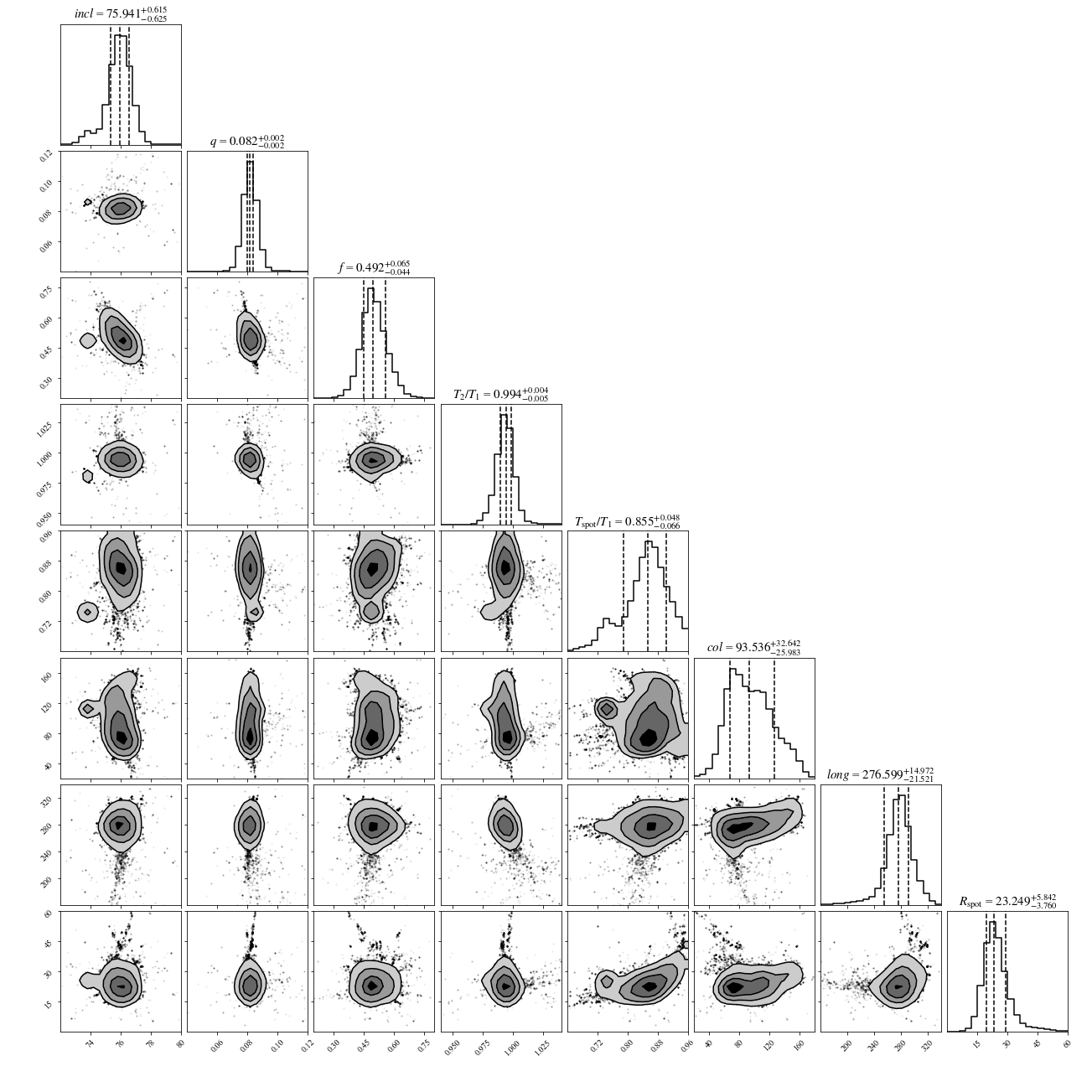}
    \caption{The MCMC corner plot for TIC 55007847. The fit model includes the orbital parameters (inclination $incl$), binary paramters (mass ratio $q$, fill-out factor $f$ and temperature ratio $T_2/T_1$) and star spot parameters (the relative temperature $T_{\rm spot}/T_1$, the colatitude of star spot $col$, the longitude of star spot $long$, and the angular radius of star spot $R_{\rm spot}$)}
    \label{fig:figure 5}
\end{figure*}

\begin{figure*}[h]
    \centering
    \includegraphics[width=0.65\textwidth]{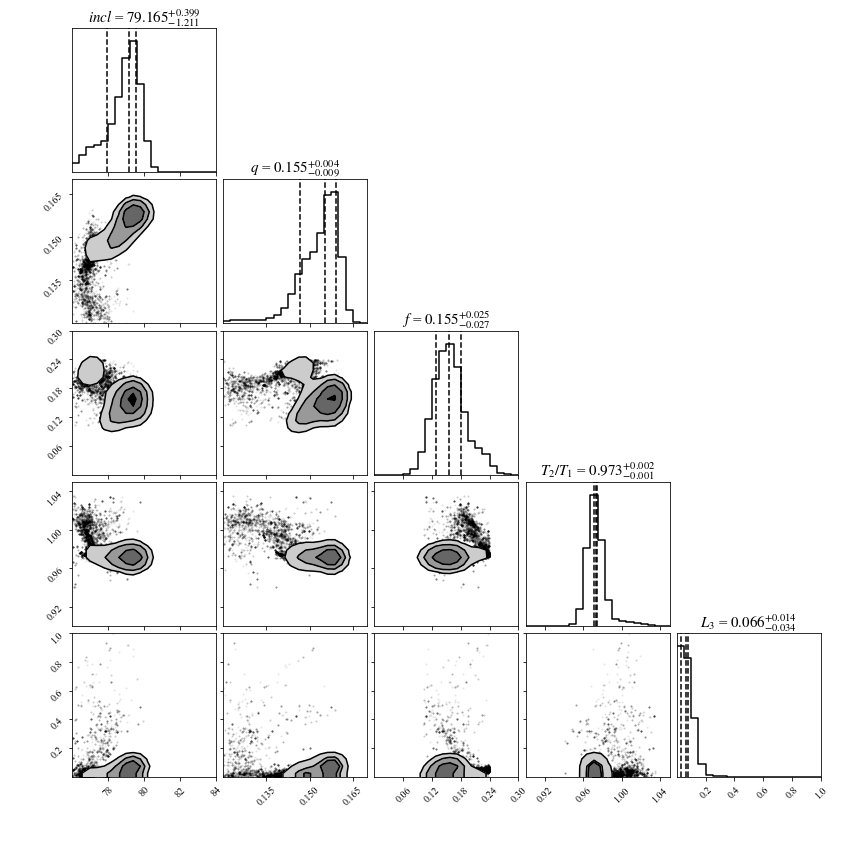}
    \caption{The MCMC corner plot with a third body of TIC 63597006, because we see a obvious periodic signal which could be caused by a third body on the O-C diagram. The fit model include the binary parameters (inclination $incl$, mass ratio $q$, fill-out factor $f$, temperature ratio $T_2/T_1$) and the third body parameters (third light fraction $L_3$)}
    \label{fig:figure 7}
\end{figure*}

\end{appendices}
\end{CJK}
\end{document}